\documentclass[prl, superscriptaddress, preprint]{revtex4-2}
\usepackage{amssymb}
\usepackage{amsfonts}
\usepackage{amsmath}
\usepackage{bm}
\usepackage{collref}
\usepackage{color}
\usepackage[markup=underlined]{changes}
\usepackage[colorlinks=true, citecolor=blue, urlcolor=blue]{hyperref}
\usepackage{dcolumn}
\usepackage{graphicx}
\usepackage{lmodern}
\usepackage{mathrsfs}
\usepackage{mathtools, slashed}
\usepackage{mathdots}
\usepackage[normalem]{ulem}
\usepackage{physics}
\usepackage{simpler-wick}
\usepackage{tikz}
\usepackage{todonotes}
\usepackage{xspace}
\makeatletter
\AddToHook{cmd/added/before}{\def\Changes@AuthorColor{purple}}
\AddToHook{cmd/deleted/before}{\def\Changes@AuthorColor{green}}
\AddToHook{cmd/replaced/before}{\def\Changes@AuthorColor{red}}
\makeatother
\setcommentmarkup{\todo[color={pink}, size=\scriptsize]{#3: #1}}

\definechangesauthor[name=lv, color=blue]{v}

\begin{document}
\title{Sp$(2N,R)$ interferometry in multi-mode Gaussian bosonic systems for optimal metrology and quantum control}

\author{Chenwei Lv}

\affiliation{State Key Laboratory of Quantum Functional Materials, Department of Physics, and Guangdong Basic Research Center of Excellence for Quantum Science, Southern University of Science and Technology, Shenzhen, China}
\affiliation{Department of Physics, The Chinese University of Hong Kong, Shatin, New Territories, Hong Kong, China}
\affiliation{The State Key Laboratory of  Quantum Information Technologies and Materials, The Chinese University of Hong Kong, Shatin, New Territories, Hong Kong, China}
\affiliation{New Cornerstone Science Laboratory, The Chinese University of Hong Kong, Shatin, New Territories, Hong Kong, China}

\author{Ren-Bao Liu}
\email{rbliu@cuhk.edu.hk}
\affiliation{Department of Physics, The Chinese University of Hong Kong, Shatin, New Territories, Hong Kong, China}
\affiliation{The State Key Laboratory of  Quantum Information Technologies and Materials, The Chinese University of Hong Kong, Shatin, New Territories, Hong Kong, China}
\affiliation{New Cornerstone Science Laboratory, The Chinese University of Hong Kong, Shatin, New Territories, Hong Kong, China}
\affiliation{Centre for Quantum Coherence, The Chinese University of Hong Kong, Shatin, New Territories, Hong Kong, China}
\date{\today}

\begin{abstract}
    Multi-mode interferometers for bosons in Gaussian states are important systems for quantum metrology with precision beyond the standard quantum limit and for bosonic quantum computing. However, there is a lack of theoretical foundation for generic $N$-mode Gaussian interferometry. In this work, we study quantum metrology and quantum control in multi-mode bosonic systems with quadratic Hamiltonians, exploiting the fundamental Sp$(2N,R)$ symmetry of such interferometers. We show that the optimal quantum control to maximize sensitivity requires aligning squeezing and displacement in the same direction. We propose Sp$(2N,R)$ echo, a multi-mode generalization of the SU$(1,1)$ interferometry, to achieve the sensitivity of phase estimation set by the quantum Fisher information. In addition, we introduce a geometrical means for reversing many-body dynamics with Sp$(2N,R)$ dynamical symmetry, such as dynamics of the bosonic Kitaev chain. Our schemes are readily realizable in optical, atomic, and mechanical platforms.
\end{abstract}
\maketitle 

Quantum effects can be exploited to push the precision of sensing beyond the standard quantum limit to the Heisenberg limit~\cite{Giovannetti2004, Giovannetti2011, Degen2017, Pezz2018}. 
After the foundation of quantum metrology~\cite{Helstrom1969}, Heisenberg-limit sensing was proposed in optical systems~\cite{Caves1981, Yurke1986}, which was realized in few-photon systems and was recently adopted to enhance the sensitivity of gravitational wave detection~\cite{Rarity1990, Nagata2007, Aasi2013}. 
Quantum-enhanced sensing has also been proposed and implemented in atomic, ionic, and spin systems, where entanglement and coherence are harnessed by spin squeezing to improve precision~\cite{Kitagawa1992, Wineland1992, Estve2008, Mao2023, Franke2023, Eckner2023}. 

A most convenient and widely adopted implementation of Heisenberg-limit quantum metrology is interferometry of bosons in Gaussian states, since such states can be readily prepared in common optical setups.
In such interferometers, the input states are Gaussian (such as coherent states and squeezed states) and the mixing elements are linear, which can be either passive (such as beam splitters for photons) or active (such as parametric down-conversion driven by an external field and an input state as an idler, which can squeeze the Gaussian distribution of the input states). 
The evolution of the state in the interferometers is generally described by a quadratic Hamiltonian (with all terms containing at most two bosonic field operators).
Gaussian states play a significant role in quantum optics and continuous-variable quantum information processing~\cite{Schumaker1986, Braunstein2005_Review, Monras2006, Pinel2012, Pinel2013, Weedbrook2012, Albert2016, Matsubara2019}. 
In cold atomic systems, Gaussian states can be accessed when the motion of the systems can be described by linearized equations~\cite{Polkovnikov2010}.
Recently, Gaussian states of optomechanical and superconducting systems have also been employed to simulate the bosonic Kitaev model~\cite{Slim2024, Clerk2024}.
Multi-mode interferometry is of particular interest for its potential in multi-parameter estimation, distributed quantum metrology~\cite{Szczykulska2016,Jacob2018,Afrnek2018,Adesso2018,Liu2019,Zhang2021,Smerzi2025} and its relevance to quantum computing such as boson sampling~\cite{Jex2017,Wang2017,Fabio2019,Zhong2020}. 
However, a general theoretical framework for multi-mode multi-boson interferometers composed of both passive and active components is still missing, which hinders the construction, assessment, and optimization of quantum metrology using such systems.

Many-body systems governed by quadratic Hamiltonians constitute an important class of quantum systems. 
Generic $N$-mode bosonic systems governed by quadratic Hamiltonians have a fundamental symmetry described by the symplectic Sp$(2N,R)$ group~\cite{Simon1994, Arvind1995, deGosson2006, Ulrik2026}. 
The noncompactness of the symplectic group naturally leads to stable/unstable transitions in dynamics. 
It also sets the ground for the Gaussian state phase space representation in quantum optics~\cite{Milburn2009}. 
The group theoretic analysis allows us to tune the parameters of the quadratic terms for coherently controlling the systems. 
It also provides us with an analytical tool to optimize the sensitivity with available resources. 

In this paper, we establish a general quantum metrology theory for $N$-mode bosonic interferometers in Gaussian states based on the Sp$(2N,R)$ symmetry. 
We introduce the Sp$(2N,R)$ echo as a multi-mode generalization of the SU$(1,1)$ interferometry~\cite{Yurke1986,Jing2011,Hudelist2014,Zhang2015,Augusto2015,Oberthaler2016,Haine2017,Lett2017,Maria2017,Lukens2018,Mao2023}.
This echo, which can invert any active multi-mode process by inserting a designated passive one, enables the design of an Sp$(2N,R)$ interferometer, where the number of bosons in a designated output mode measures the applied phase with Heisenberg-limit sensitivity.
The Sp$(2N,R)$ interferometry can also be employed to reverse quantum dynamics without changing the sign of the Hamiltonian, providing a useful tool for coherent quantum control. 

We consider a general quadratic Hamiltonian of $N$ bosonic modes, 
\begin{equation}
    \hat H =\frac{1}{2}h_{ab} {\hat\xi}^a {\hat\xi}^b, ~\label{Eq:QuadraticH}
\end{equation}
where $h$ is a $2N\times 2N$ symmetric real matrix and ${\hat\xi}=(\hat q_1,\dots,\hat q_N,\hat p_1,\dots,\hat p_N)^{\rm T}$, $\hat q_j = (\hat a_j^\dag+\hat a_j)/\sqrt{2}$, $\hat p_j = i(\hat a_j^\dag-\hat a_j)/\sqrt{2}$ are the quadratures with
$\hat a_j(\hat a_j^\dag)$ being the annihilation (creation) operator of the $j$-th bosonic mode.
We adopt the Einstein summation convention. 
Symbols with (without) hat represent operators acting on the infinite-dimensional multi-mode Hilbert space (finite-dimensional matrices or vectors).
To show the Sp$(2N,R)$ symmetry, we write the Heisenberg equations of motion for ${\hat\xi}$,
\begin{equation}
    \partial_t {\hat\xi} = \Omega h {\hat\xi},
\end{equation}
where
\begin{equation}
    \Omega = \begin{pmatrix}
        0 & {\bf I}_{N} \\
        -{\bf I}_{N} & 0 \\
    \end{pmatrix}, \notag
\end{equation}
with ${\bf I}_{N}$ denoting the $N\times N$ identity matrix. 
The matrix $U(t,0)=\mathcal{T} e^{\int_0^{t}\Omega h(t') dt'}$ that propagates ${\hat\xi}(0)$ to ${\hat\xi}(t)$ is an element of the real symplectic group Sp$(2N,R)$ satisfying $U\Omega U^{\rm T} = \Omega$. 
Therefore, any dynamics generated by a time-dependent quadratic bosonic Hamiltonian {$\hat{\mathcal U}(t,0)=\mathcal{T} e^{-i\int_0^{t} \hat H(t') dt'}$} belongs to the Sp$(2N,R)$ group, with $\hat{\mathcal{U}}^\dag{\hat\xi} \hat{\mathcal{U}} = U{\hat\xi}$. 

To optimize the sensitivity of probe states generated by a quadratic Hamiltonian, we compute the quantum Fisher information (QFI) for phase estimation. 
Without loss of generality, the parameter to be measured $\varphi$ is encoded to the probe state $\ket{\Psi}$ through $\ket{\Psi(\varphi)}=e^{-i\varphi\hat P}\ket{\Psi}$, where ${\hat P}\equiv \sum_{j=1}^{N}\hat a^\dag_j \hat a_j= \left({\hat\xi}^{\rm T}{\hat\xi}-N\right)/2$ is the total number operator.
QFI $F_\varphi$ sets the ultimate uncertainty of estimating $\varphi$ through the quantum Cram\'er Rao bound (QCRB)~\cite{Helstrom1969},
\begin{equation}
\delta \tilde\varphi \ge  F_\varphi^{-1/2}, \label{Eq:QCramerRao}
\end{equation}
where $\tilde\varphi$ is an unbiased estimator of $\varphi$.
We assume $\ket{\Psi}$ is a pure Gaussian state generated by applying $\hat{\mathcal U}$ to the multi-mode vacuum $\ket{0}$.
If the initial reference state is pure Gaussian but not the vacuum, it can be absorbed into a redefinition $\hat{\mathcal U}$.
In this case, the QFI for phase estimation is proportional to the quantum fluctuations of the number operator {$\hat{P}$}~\cite{Smerzi2014, Liu2019}.
In the Heisenberg picture, $\hat{P}$ transforms into $\left({\hat\xi}^{\rm T}U^{\rm T}U{\hat\xi}-N\right)/2$. 
Using Wick's theorem for $\left\langle {\hat\xi}^{i}{\hat\xi}^{j}\cdots {\hat\xi}^{l}\right\rangle$~\cite{Peskin2018} and $\wick{\c2 {\hat\xi^m}\c2{\hat\xi^n}}=\frac{1}{2}(\delta^{mn}+i\Omega^{mn})$, we find
\begin{equation}
    F_\varphi = 4\left(\left\langle \hat{P}^2\right\rangle -\left\langle \hat{P}\right\rangle^2\right)=\frac{1}{2}\Tr\left[\left(UU^{\rm T}\right)^2\right]-N.~\label{Eq:Fisher}
\end{equation}

The optimization problem can be formulated as maximizing the QFI with a given cost of $\hat H$.
We adopt the method of continuous circuit complexity~\cite{Nielsen2006} and define the cost function ${\mathcal C}$ of the Hamiltonian $\hat H$ through the parameters of its quadratic terms,
\begin{equation}
    {\mathcal C}(h)=\sqrt{\sum_{i,j}p_{ij}h^2_{ij}},\label{Eq:cost}
\end{equation}
where $p_{ij}$ are the penalty factors.
${\mathcal C}(h)$ quantifies the required resource of realizing $\hat H$.
It induces a metric on the space of pure Gaussian states.
Unlike the quantum metric whose line element amounts to QFI for pure states~\cite{Provost1980, Caves1994}, the trajectory length, $C=\int_0^{t} {\mathcal C}(h(t'))dt'$, measures the difficulty of generating the probe state along the path $\hat {\mathcal{U}}(t',0)\ket{0}$.
In the following, we will take $p_{ij}=1$ for simplicity, where ${\mathcal C}(h)$ becomes the Frobenius norm ${\mathcal C}(h) = ||h||=\sqrt{\Tr(h^2)}$.
The analytical approach also applies to general $p_{ij}$.
The time derivative of QFI is 
\begin{equation}
    \frac{dF_\varphi}{dt} \equiv \dot F_\varphi = \Tr([\Omega,h](U U^{\rm T})^2).~\label{Eq:FisherChange}
\end{equation}
It is useful to expand $h = \sum_{m=0,x,y,z} \sigma_m\otimes h_m=h_1+h_2$,
where $h_1=\sigma_0\otimes h_0+\sigma_y\otimes h_y$, $h_2=\sigma_x\otimes h_x+\sigma_z\otimes h_z$, with $\sigma_0$ and $\sigma_{x,y,z}$ being $2\times 2$ identity matrix and Pauli matrices, respectively, and $h_{0,x,z}$ ($ih_y$) being real and symmetric (anti-symmetric) $N\times N$ matrices.
From this decomposition, we have $||h||^2 = ||h_1||^2+||h_2||^2$. 
Note that $h_1$ generates the $U(N)$ subgroup of Sp$(2N,R)$, which conserves the total boson number and does not change $F_\varphi$. 
We set $h_1=0$ since $[\Omega, h_1]=0$.
On the other hand, $h_2$ only contains pair creation/annihilation terms and can be regarded as an interaction Hamiltonian for parametric down-conversion. 
Dynamics generated by $h_2$ is unstable, leading to QFI exponentially increasing with time. 
For instance, when $h=h_2$ and is time-independent, $\Omega h=-h\Omega=(\Omega h)^{\rm T}$. 
We directly find $F_\varphi=\Tr(e^{4\Omega h t})/2-N \sim \langle \hat P\rangle^2=(\Tr(e^{2\Omega h t})/2-N)^2/4$, which reaches the Heisenberg limit of quantum sensing.

\begin{figure}
    \centering
    \includegraphics[width=0.76\textwidth]{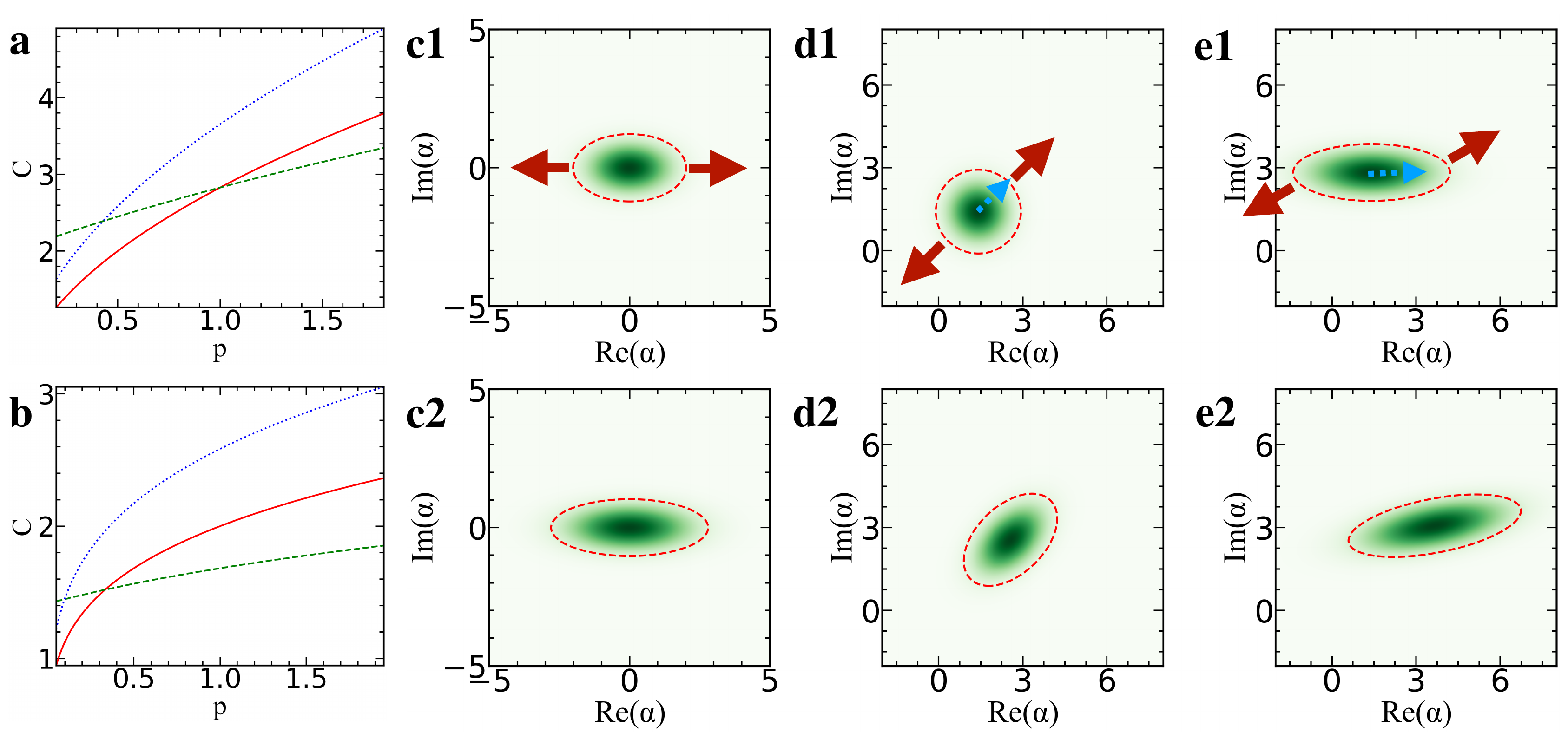}
    \caption{
 (a,b) Total cost $C$ of generating probe states with the same QFI from the vacuum for $N=2$ as a function of anisotropy $p$ of the penalty factor, $p_{i\neq j}=1$, $p_{ii}=p$, with $\hat H=\hat{a}_1^{\dag 2}+{\rm h.c.}$ (red solid curve), $\hat H=(\hat{a}_1^{\dag}+\hat{a}_2^{\dag})^2/2+{\rm h.c.}$ (green dashed curve) and $\hat H=(\hat{a}_1^{\dag 2}+\hat{a}_2^{\dag 2})/2+{\rm h.c.}$ (blue dotted curve) applied for $t=1,1,1.827$, respectively. 
 A quartic cost function $\mathcal{C}_4(h)=(\sum p_{ij}h^4_{ij})^{1/4}$ is considered in (b).
 (c-e) Husimi-Q representation $Q(\alpha)$ for squeezed, displaced, and displaced-squeezed probe states before (c1, d1, e1) and after (c2, d2, e2) applying the optimized Hamiltonians that maximize the QFI growth. Red solid (blue dotted) arrows denote the squeezing (displacement) direction of the optimized Hamiltonian.}
    \label{Fig:optimal}
\end{figure}

To optimize $h_2$ for QFI, we adopt Euler/Bloch-Messiah decomposition of the Sp$(2N,R)$ group~\cite{Arvind1995, Braunstein2005_BM, Houde2024},
\begin{equation}
    U = S_1e^{\Lambda} S_2,\quad \Lambda =\sigma_z\otimes \lambda,~\label{Eq:Euler}
\end{equation}
where $\lambda={\rm diag}(\lambda_1,\dots,\lambda_N)$ is a diagonal matrix with $\lambda_1\geq \lambda_2\geq \dots\geq \lambda_N\geq 0$.
$S_{1,2}\in$ Sp$(2N,R)\cap$ SO$(2N)\simeq$ U$(N)$ are orthogonal and symplectic. 
We note that the quadrature basis is related to the creation/annihilation operator basis by the transformation $g{\hat\xi} =(\hat a_1,\dots,\hat a_N,\hat a_1^\dag,\dots \hat a_N^\dag)^{\rm T}$, where
\begin{equation}
   g = \frac{1}{\sqrt{2}}\begin{pmatrix}
       {\bf I}_N&i{\bf I}_N\\
       {\bf I}_N&-i{\bf I}_N
   \end{pmatrix}.~\label{Eq:gTransform}
\end{equation}
Using Eq.~\eqref{Eq:Euler} and changing the basis from quadratures to creation/annihilation operators, Eq.~\eqref{Eq:FisherChange} becomes 
\begin{equation}
    \dot F_\varphi = 4\Re \left[ \Tr\left(u_1(h_x-ih_z)u_1^{\rm T}\sinh(4\lambda)\right)\right],
\end{equation}
where $S_1={\rm diag}(u_1,u_1^*)$ in the new basis, and $u_1$ is unitary.  
$h_x-ih_z$ is complex symmetric and can always be diagonalized by a unitary matrix $u$ through the Autonne-Takagi factorization, $h_x-ih_z=uBu^{\rm T}$, where $B={\rm diag}(b_1,\cdots,b_N)$ is a non-negative diagonal matrix and $||h_2||^2=2||B||^2$. 
Physically, $u_1$ and $u^\dag$ represent directions of the supermodes squeezed in the probe state and by the interaction Hamiltonian, respectively, while $\lambda_i$ and $b_i$ are the corresponding squeezing strengths. 
$\dot F_\varphi = 4\Tr(B\sinh(4\lambda))$ if we choose $u^\dag=u_1$. 
It is maximized for a given probe state when $B=||h||\sinh(4\lambda)[2{\rm Tr}(\sinh^2(4\lambda))]^{-1/2}$. 
The maximum growth rate is $(\dot F_\varphi)_{\rm max} = 2\sqrt{2}||h||[{\rm Tr}(\sinh^2(4\lambda))]^{1/2}$.
When only one supermode is squeezed, $\lambda_{j>1}=0$, $(\dot F_\varphi)_{\rm max} = 2\sqrt{2}||h||\sinh(4\lambda_1)$, the QFI increases most rapidly when the probe state is squeezed along its maximally squeezed direction.
This is reminiscent of the single-mode squeezed vacuum as the optimal Gaussian state for quantum sensing~\cite{Matsubara2019}.
The relation $(\dot F_\varphi)_{\rm max}\propto ||h||$ makes the multi-mode approach advantageous over single-mode approaches, as $h$ is a $2N\times 2N$ matrix such that a larger $||h||$ could be more accessible in many-body systems.
In Fig.~\ref{Fig:optimal}(a), we show the total cost of generating a probe state from vacuum under anisotropic penalty factors, with $p_{i\neq j}=1$ and $p_{ii}=p$. 
While the generated probe states have the same QFI, the probe state squeezed along the physical/collective mode direction has a lower/higher cost when $p<1$.
This condition is reversed when $p>1$, meaning that squeezing a collective supermode becomes more advantageous than squeezing individual physical modes when taking the multi-mode approach.
We also note that here we adopt $h_{ij}^2$ in Eq.~\eqref{Eq:cost}. 
Including higher-order terms such as $h_{ij}^4$ would also favor collective supermodes, as shown in Fig.~\ref{Fig:optimal}(b), and the resulting geometry of Gaussian states becomes Finslerian instead of Riemannian~\cite{Nielsen2005,Pfeifer2019}.

Our results also apply when the probe state is a displaced Gaussian state, obtained by applying $\hat{\mathcal{U}}$ to a multi-mode standard coherent state displaced from the vacuum $\ket{0}$ by $\hat{\mathcal{D}}$. 
In this case, we add linear terms in the Hamiltonian Eq.~\eqref{Eq:QuadraticH},
$\hat H'=h_{ab}{\hat\xi}^a{\hat\xi}^b/2+\bar f_a{\hat\xi}^a$, such that $\hat{\mathcal{U}}'=\mathcal{T}e^{-i\int_0^t\hat H'(t')dt'}=\hat{\mathcal{U}}\hat{\mathcal{D}}$.
To optimize the QFI, we extend the column vector of quadrature operators by one more dimension from ${\hat\xi}$ to ${\hat\xi}'=({\hat\xi}^{\rm T},\hat{\mathbb{I}})^{\rm T}$, where $\hat{\mathbb{I}}$ is the many-body identity operator. 
The Heisenberg equation of motion for ${\hat\xi}'$ is
\begin{equation}
    \partial_t {\hat\xi}' = \begin{pmatrix}
        \Omega h & \Omega \bar f \\ 
        0 & 0 \\ 
    \end{pmatrix}{\hat\xi}' \equiv s{\hat\xi}'.
\end{equation}
The propagator becomes 
\begin{equation}
    U' = \mathcal{T}e^{\int_0^t s(t') dt'}=\begin{pmatrix}
        U & f \\ 
        0 & 1 \\ 
    \end{pmatrix},
\end{equation}
where $f$ denotes the displacements and its explicit form is given by the Dyson series of integrating $\partial_t f = \Omega h f + \Omega \bar f$. 
$U'$ is a group element of the affine symplectic group ISp$(2N,R)$~\cite{deGosson2006}. 
The QFI is then modified to $F_\varphi' = F_\varphi+2f^{\rm T}(UU^{\rm T})f$.
Again, we take the time derivative for optimizing $\hat H'(t)$ that maximizes QFI growth 
\begin{equation}
    \dot{F}'_\varphi= \dot{F}_\varphi
        +2f^{\rm T} \left\{U U^{\rm T}, [\Omega, h]\right\} f
        +4f^{\rm T} U U^{\rm T} \Omega \bar f. \label{Eq:FisherChange_2}
\end{equation}
Now, the cost function induces a metric on the space of pure displaced Gaussian states. 
The difficulty of generating a displacement is quantified by the cost of linear terms, where we define as $||\bar f||=\sqrt{\bar f^a\bar f_a}$, the cost function becomes $\mathcal{C}'(h,\bar f)=||h||+||\bar f||$.
For a probe state with given $U$ and $f$, the row vector $4f^{\rm T} U U^{\rm T} \Omega$ is fixed.
Since $||\bar f||$ is rotationally invariant, the QFI growth is optimized when $\bar f\propto -\Omega UU^{\rm T}f$. 
Using Eq.~\eqref{Eq:Euler}, the induced displacement $\Omega \bar f$ is along $S_1e^{2\Lambda}S_1^{\rm T}f$, which moves the probe state towards the maximal squeezing direction.
In the absence of squeezing in the probe state, $U={\bf I}_{2N}$, $\bar f\propto -\Omega f$, $\bar f_a{\hat\xi}^a$ generates displacement along $f$ and the first term $\dot{F}_\varphi$ in $\dot{F}'_\varphi$ vanishes. 
The second term becomes $4f^{\rm T} [\Omega, h] f$, which is optimized when $h$ squeezes the probe state along $f$.
To show this, we define $f=(f^{\rm T}_q,f^{\rm T}_p)^{\rm T}$ and $f'=f_q+if_p$. 
Using the transformation in Eq.~\eqref{Eq:gTransform}, we find $f^{\rm T} [\Omega, h] f=2{\rm Re}\left[f^{\prime T}(h_x+ih_z)f'\right]$, which is optimized when $h_x+ih_z \propto f^{\prime *}f^{\prime \dag}$.
When the Gaussian state is squeezed and displaced, the first and second terms compete and the optimal $h$ needs to be computed numerically.

In Fig.~\ref{Fig:optimal}, we show the Husimi-Q representation $Q(\alpha) = |\langle\alpha|\psi\rangle|^2/\pi$ of a probe state before and after applying the optimal Hamiltonian for a short time~\cite{Crispin2004}. 
For simplicity, we use a single mode as an example. 
$|\alpha\rangle=e^{-|\alpha|^2/2}\sum_{n=0}^{\infty}\alpha^n|n\rangle/\sqrt{n!}$.
When there is no displacement in the initial probe state, the optimal squeezing direction aligns with the original maximal squeezing direction, as depicted in Fig.~\ref{Fig:optimal}(c1,c2).
In Fig.~\ref{Fig:optimal}(d1,d2), we show that both the optimal displacement and squeezing directions coincide with the original displacement direction in the absence of squeezing in the probe state. 
In Fig.~\ref{Fig:optimal}(e1,e2), where the initial probe state is both squeezed and displaced, the optimal displacement direction is $UU^{\rm T}f$, while the squeezing direction is determined numerically. 
We find that the displacement and squeezing directions become more closely aligned after applying the optimal Hamiltonian.

For an interferometer with unitary dynamics and vacuum at the output port in the absence of signal, performing projective measurements of $\hat{P}$ at the output port is optimal and saturates the QCRB (Supplementary material).
For this purpose, we introduce an Sp$(2N,R)$ echo, where echo pulses are inserted into the dynamics and lead to a revival of the initial state.
To formulate the Sp$(2N,R)$ echo and Sp$(2N,R)$ interferometer, we consider Hamiltonians that generate the active and passive processes $\hat H=\frac{1}{2}h_{ab}{\hat\xi}^a{\hat\xi}^b$ and $\hat H^{\rm e} = \frac{1}{2}h^{\rm e}_{ab}{\hat\xi}^{a}{\hat\xi}^{b}$.
Here, the echo Hamiltonian $\hat H^{\rm e}$ conserves the boson number and belongs to the $\mathfrak{u}(N)$ subalgebra. 
In the Sp$(2N,R)$ echo, $\hat H$ is applied for a duration $T_0$, then $\hat H^{\rm e}$ is applied for time $T_1$, and $\hat H$ is applied again before the measurement at time $T_1+2T_0$.
The dynamics $U$ generated by $\hat H$ can be reversed by inserting a U$(N)$ subgroup element $V$, such that 
\begin{equation}
    (VUVU)^2={\bf I}_{2N}, \quad U=e^{\Omega h T_0}.~\label{Eq:EchoSequence}
\end{equation}
Instead of reversing all signs in $\hat H$, $V$ can be realized by beam-splitters and phase shifters. 
With the Euler decomposition Eq.~\eqref{Eq:Euler}, Eq.~\eqref{Eq:EchoSequence} becomes $(VS_1e^{\Lambda} S_2 V S_1e^{\Lambda} S_2 )^2={\bf I}_{2N}$.
$V = S_2^{\rm T} \Omega S_1^{\rm T}=e^{\Omega h^{\rm e} T_1}$ is an element of $U(N)$ and makes this equality hold, where $e^{\Lambda}\Omega e^{\Lambda} = \Omega$ is used. 
The corresponding echo Hamiltonian has
\begin{equation}
    h^{\rm e} = -\frac{1}{T_1}\Omega\log(U^{\rm T}(UU^{\rm T})^{-1/2}\Omega).~\label{Eq:EchoH}
\end{equation}
We note $UVUV=-{\bf I}_{2N}$. 
$U^{\rm T}(UU^{\rm T})^{-1/2}$ is unique and is the transpose of the orthogonal part of the polar decomposition of $U$. 
$h^{\rm e}$ is unique up to {$2\pi n/T_1$ in the frequencies of its eigenmodes, $n\in\mathbb{Z}$}, with ambiguity arising from the logarithm taken on different branches.

The SU$(1,1)$ interferometry is a special case of the Sp$(2N,R)$ echo. 
When $N=1$, $h=\sigma_x$, $V=\Omega$ and $\hat H^{\rm e}=\frac{1}{2}(\hat p^2+\hat q^2)$ for $T_1 = \pi/2$, the SU(1,1) interferometry and $\hat H^{\rm e}$ can be created by phase shifters~\cite{Yurke1986, Chen2020}. 
When $h$ is a general $2\times 2$ real symmetric matrix, $\hat H^{\rm e}$ can still be solved analytically, which creates an SU$(1,1)$ echo~\cite{Lyu2020, Lv2020}. 
Follow-up works demonstrate a quantum echo in reversing dynamics of two bosonic modes~\cite{Wang2024}. 
Here in Eq.~\eqref{Eq:EchoH}, we provide a general formula for the echo Hamiltonian such that any quadratic Hamiltonian dynamics in an arbitrary number of modes can be reversed. 

Since $\hat{\mathcal{V}}=e^{-i\hat H^{\rm e}T_1}$ leaves the vacuum invariant, $\hat{\mathcal{U}}\hat{\mathcal{V}}\hat{\mathcal{U}}\ket{0}\propto\ket{0}$, the sequence $UVU$ forms the Sp$(2N,R)$ echo interferometer.
If a phase shift is applied before or after the echo pulse $V$, the final state is no longer the vacuum.
The QFI of the state at the end of the echo sequence with respect to the phase shift $\varphi$ is $F_\varphi=4\lim\nolimits_{\varphi\to 0}[1-L(\varphi)]/\varphi^2$,
where $L(\varphi)=|\langle 0|\hat {\mathcal{U}}\hat {\mathcal{V}}e^{-i\varphi\hat{P}}\hat {\mathcal{U}}|0\rangle|^2$ is the Loschmidt echo~\cite{Wisniacki2012, Tommaso2016}. 
A short calculation shows that it is the same as the QFI in Eq.~\eqref{Eq:Fisher}.
We also compute the boson number at the output port in the small $\varphi$ limit, which reads (Supplementary material)
\begin{equation}
    \langle\hat{P}(\varphi)\rangle = \frac{1}{2}\varphi^2 F_\varphi.
    \label{Eq:N_varphi}
\end{equation}
The small $\varphi$ signal is amplified by $\sqrt{F_\varphi}$ at the final boson detection port, and can be measured by detecting the total output boson number.

\begin{figure}
    \centering
    \includegraphics[width=0.48\textwidth]{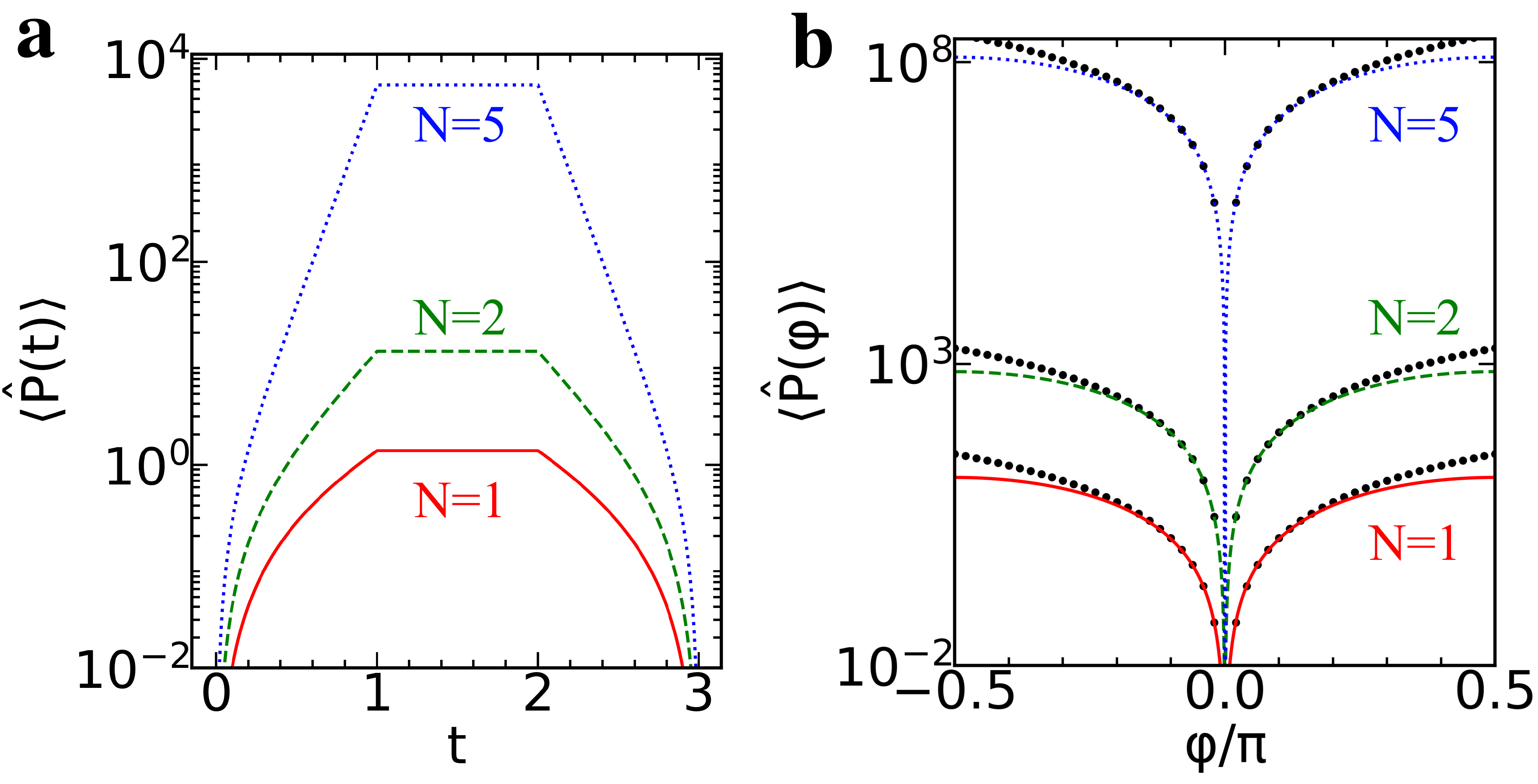}
    \caption{
      Total boson number $\langle\hat{P}\rangle$ 
      (a) during the echo sequences with no additional phase applied 
      (b) at the final detection port {$t=T_1+2T_0$} with an additional phase shift $e^{-i\varphi \hat{P}}$ applied before the echo Hamiltonian.
      Results for $N=1,2,5$ are represented by the red solid, green dashed and blue dotted curves, respectively.
      Black dots are analytical results at small $\varphi$ in Eq.~\eqref{Eq:N_varphi}.}
    \label{Fig:echo}
\end{figure}

To demonstrate our results, we let the quantum state be squeezed along the collective direction $x=(1,1,\dots,1)^{\rm T}$.
The corresponding Hamiltonian has $h = -\sigma_z\otimes {\bf O}_N$, where ${\bf O}_N$ denotes an $N\times N$ matrix with all entries being $1$.
Its QFI is $F_\varphi = 2\sinh^2(2NT_0)$. Using Eq.~\eqref{Eq:EchoH}, we find $h^{\rm e}={\bf I}_{2N}\pi/2$ for $T_1=1$.
In Fig.~\ref{Fig:echo}(a), we show the total boson number as a function of time with $T_0=1$ and $N=1,2,5$.
In Fig.~\ref{Fig:echo}(b), an additional phase shift $e^{-i\varphi\hat{P}}$ is added before the echo pulse, which effectively changes $T_1$. 
This leads to an imperfect echo and $\langle\hat{P}(\varphi)\rangle$ at $T_1+2T_0$ becomes finite. 
For small $\varphi$, $\langle\hat{P}(\varphi)\rangle$ follows the analytical result in Eq.~\eqref{Eq:N_varphi}. 
An exponentially larger sensitivity for measuring $\varphi$ is achieved for bigger $N$ as $\mathcal{C}=||h||=\sqrt{2}N$ increases linearly, providing a larger exponential growth rate of the QFI.
We emphasize that, although $x$ is a supermode obtainable through a unitary transformation from a single mode, it enables the distribution of single-mode squeezing across many modes, and therefore $||h||$ is enhanced~\cite{Fabre2020}.
Especially, if a quartic cost function $\mathcal{C}_4=(\sum h_{ij}^4)^{1/4}$ is applied, $F_\varphi=2\sinh^2(\sqrt{N}2^{3/4}C)$. 
The QFI increases subexponentially as $N$ increases when the total cost $C=\mathcal{C}_4T_0$ is fixed. 
This demonstrates an important advantage of multi-mode interferometers versus two-mode ones.

\begin{figure}
    \centering
    \includegraphics[width=0.48\textwidth]{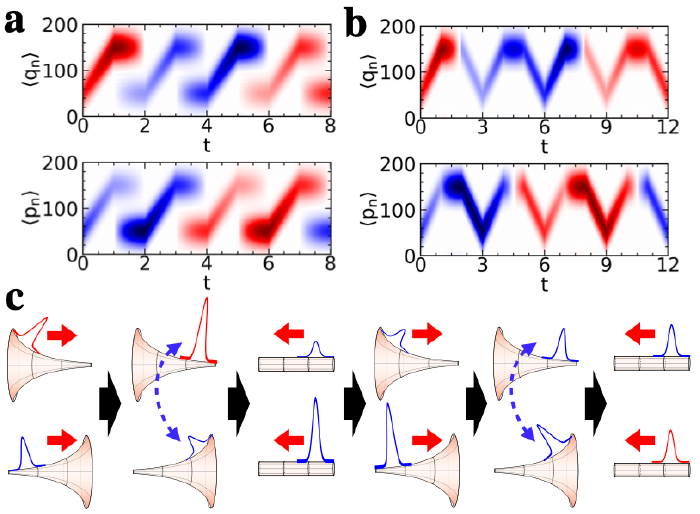}
    \caption{ Sp$(2N,R)$ echo for reversing quantum dynamics of the BKC model. 
    (a) Numerical simulation of the wavepacket dynamics in quadratures during the echo sequence. We choose $\Delta=0.4$ and $J=50$. 
    The initial Gaussian wavepacket is located at $n_0=50$ with $\sigma=20$. $T_0=T_1=1$, the wavepacket returns to its initial position at $t=8$.
    (b) Numerical simulation of the wavepacket dynamics with the same parameters but a two-step driving sequence to deliver the echo pulse. 
    $T_0=1$, the state revives at $t=12$.
    (c) A schematic plot of the first half of the echo sequence in (b) from $t=0$ to $t=6$ based on the underlying curved space dynamics. 
    The dimension reduction is performed by choosing PBC in the azimuthal direction of the hyperbolic surfaces and flat cylinders. 
    Blue dashed arrows (red arrows) represent the switch (moving directions) of the wavepackets in the space of different quadratures.} 
    \label{Fig:reverse}
\end{figure}

In addition to the applications in interferometry, our results directly apply to the reversal of quantum dynamics in many-body bosonic systems. 
As an example, we consider the bosonic Kitaev chain (BKC) with periodic boundary conditions (PBC). 
As a bosonic analog of the celebrated fermionic Kitaev model, it exhibits a quadrature-dependent chiral transport, curved space dynamics, and entanglement phase transitions~\cite{Clerk2018, LV2024, Lee2024}.
The Hamiltonian of BKC is
\begin{equation}
    \hat H_{\rm BK} = - i\sum_{j=1}^{N} \left(\Delta \hat a_j \hat a_{j+1} + J \hat a_j^\dag \hat a_{j+1} \right)+ {\rm h.c.},~\label{Eq:BKC}
\end{equation}
where $\Delta$ and $J$ are the amplitudes of the nearest neighbor pair creation/annihilation and tunneling terms, respectively.
In terms of quadratures, $h_{\rm BK}=\sigma_+ \otimes h_{\rm HN} + \sigma_- \otimes h_{\rm HN}^{\rm T}$, where 
$(h_{\rm HN})_{ij}=\delta_{i+1,j}(\Delta+J)+\delta_{i,j+1}(\Delta-J)$ is the Hamiltonian of Hatano-Nelson model, and $\sigma_\pm = (\sigma_x\pm i\sigma_y)/2$. 
The Hatano-Nelson model is a well-known non-Hermitian model with nonreciprocal tunneling, which leads to chiral transport~\cite{Hatano1996}. 
The wavepacket is amplified (attenuated) when it travels to the direction with a greater (smaller) tunneling strength.
$\Omega h_{\rm BK} = (\sigma_z-\sigma_0)/2 \otimes h_{\rm HN}+(\sigma_z+\sigma_0)/2 \otimes h_{\rm HN}^{\rm T}$ is block diagonal in the $q$ and $p$ quadratures with opposite chirality, which results in the quadrature-dependent chiral transport.

The echo Hamiltonian can be directly constructed from Eq.~\eqref{Eq:EchoH}. 
In Fig.~\ref{Fig:reverse}(a), we set the initial state as a coherent state with a Gaussian distribution of both quadratures, $\langle q_n\rangle = e^{-(n-n_0)^2/2\sigma^2}$ and $\langle p_n\rangle = -e^{-(n-n_0)^2/2\sigma^2}$. We choose $T_0=T_1=1$, such that the full echo sequence in Eq.~\eqref{Eq:EchoSequence} is completed at $t=8$.
Since $UVUV=-{\bf I}_{2N}$, the wavepacket returns to its initial position with an opposite sign at $t=4$.

This designed echo sequence admits a geometrical explanation when we use a two-step driving protocol to deliver the echo pulse. 
Notice the echo pulse in Eq.~\eqref{Eq:EchoH} contains $U^{\rm T}(UU^{\rm T})^{-1/2}$ and $\Omega$, we observe that $U^{\rm T}(UU^{\rm T})^{-1/2}$ is generated by applying
\begin{equation}
    \hat H^{\rm e}_{\rm BK} = \sum_{j=1}^{N} (iJ \hat a_j^\dag \hat a_{j+1} + {\rm h.c.}),
\end{equation}
for a duration $T_0$, and $\Omega$ is generated by applying
\begin{equation}
    \hat H_{\Omega} = \frac{\pi/2}{T_0}\sum_{j=1}^{N} \hat a_j^\dag \hat a_{j}=\frac{\pi/2}{T_0}\hat{P},
\end{equation}
for a duration $T_0$, such that $e^{+i\hat H_{\Omega}T_0}{\hat\xi} e^{-i\hat H_{\Omega}T_0}=\Omega{\hat\xi}$.
The echo pulse is then created by applying $\hat H_{\Omega}$ followed by $\hat H^{\rm e}_{\rm BK}$, each for a duration of $T_0$. 
As shown in Fig.~\ref{Fig:reverse}(b), the initial state revives at the end of the echo sequence at $t=12$, where we choose $T_0=1$.

This process can be understood through its duality to curved space dynamics.
The effective theory of $h_{\rm HN}$ at PBC for a slowly varying $\psi_n$ is given by 
\begin{equation}
    i\partial_t \psi = iv\bigg(\partial_s+\frac{\Delta}{Jd}\bigg) \psi,~\label{Eq:curved}
\end{equation}
where $s=nd$ denotes the spatial coordinate with $d$ being the lattice constant and the velocity $v=2Jd$.
This is a projection of a 2+1 dimensional relativistic Schr\"odinger equation on a hyperbolic space with the metric tensor ${\bf g}_\pm = ds^2+e^{\pm 2\sqrt{\kappa} s}dx^2$.
The momentum along $x$-direction is set to zero, $+$ ($-$) sign is taken for $\Delta/J>0$ ($<0$), and the curvature is $-\kappa=-4\Delta^2/(J^2d^2)$.
Eq.~\eqref{Eq:curved} has a wavepacket solution $\psi(s, t)=f(s+vt)e^{\mp \sqrt{\kappa}s/2}$, such that the quantity $C = \int |\psi|^2\sqrt{g_\pm}ds$ is conserved with $g_\pm={\rm det}{\bf g}_\pm=e^{\pm 2\sqrt{\kappa}s}$.
Any wavepacket will propagate dispersionlessly with velocity $-v$ with an exponential amplification or attenuation.
When $\Delta=0$, the hyperbolic space becomes flat. 

The curved spaces with metric ${\bf g}_\pm$ take funnel shapes with opposite opening directions, known as pseudospheres, are illustrated in Fig.~\ref{Fig:reverse}(c).
During the time interval $0<t<1$, the two wavepackets in the $p$ and $q$ quadratures propagate with the same velocity $v$ on these two pseudospheres. 
Their amplitudes are exponentially amplified or attenuated due to the conservation of $C$. 
During $1<t<2$, the wavepacket in the $p$ ($q$) quadrature is transported to the $q$ ($p$) quadrature, where the pseudosphere has an opposite opening. 
Subsequently, their amplitudes are exponentially attenuated or amplified during $3<t<4$, respectively, returning to their initial values. 
$\hat H^{\rm e}_{\rm BK}$ in the time interval $2<t<3$ and $5<t<6$ pulls the wavepackets backward with velocity $-v$ in a flat space to their initial positions.
During $4<t<5$, the wavepackets in the $p$ and $q$ quadratures are switched back, with an additional $-$ sign. 
A schematic plot based on the underlying relativistic wavepacket dynamics in the curved space is illustrated in Fig.~\ref{Fig:reverse}(c), which provides a geometrical interpretation of the echo sequence.

Using the Sp$(2N,R)$ algebraic method and continuous circuit complexity, we have demonstrated the optimization of the bosonic quadratic Hamiltonian that maximizes the QFI growth with the cost defined by the Frobenius norm of its coefficient matrix. 
After applying the optimized Hamiltonian, the Gaussian state is squeezed or displaced along the maximum squeezing or displacement direction of the original ones.
Such a geometric approach works for a general cost function. 
We show that an anisotropic cost function containing higher-order terms could penalize the concentration of energy in few modes, making the multi-mode approach advantageous over single-mode ones.
We also propose the Sp$(2N,R)$ echo, which can be used to construct an Sp$(2N,R)$ interferometer. 
This allows us to detect the applied phase shift through the total boson number, achieving the sensitivity set by the QFI. 
Additionally, the Sp$(2N,R)$ echo can reverse quantum dynamics in bosonic systems with quadratic Hamiltonians, which still applies when linear terms are included (Supplementary material).

Given the extensive experimental studies of the quadratic bosonic systems on various platforms, it is promising that our theoretical predictions can soon be tested in laboratory settings. 
While our method is algebraic, it can be applied to various physical systems with Sp$(2N,R)$ dynamical symmetry or generalized to other symmetry groups, such as fermionic Gaussian states.
As a geometric tool for manipulating many-body dynamics, the Sp$(2N,R)$ echoes are useful in quantum information processing, metrology, and quantum control.
We hope our work will stimulate more studies in coherent control of quantum dynamics using group-theoretic and geometric methods.

\acknowledgements{
This work was supported by the Innovation Program for Quantum Science and Technology Project No. 2023ZD0300600, the Hong Kong Research Grants Council - General Research Fund (Project No. 14302121), the National Natural Science Foundation of China/Hong Kong Research Council Collaborative Research Scheme Project CRS-CUHK401/22, the Hong Kong Research Grants Council Senior Research Fellow Scheme Project SRFS2223-4S01, and the New Cornerstone Science Foundation.
}

\onecolumngrid
\newpage
\vspace{0.4in}
\centerline{\bf\large Supplementary Materials for }
\centerline{\bf\large ``Sp$(2N,R)$ interferometry in multi-mode Gaussian bosonic systems}
\centerline{\bf\large for optimal metrology and quantum control''}
\setcounter{equation}{0}
\setcounter{figure}{0}
\setcounter{table}{0}
\makeatletter
\renewcommand{\theequation}{S\arabic{equation}}
\renewcommand{\thefigure}{S\arabic{figure}}
\renewcommand{\thetable}{S\arabic{table}}
\vspace{0.2in}

\section{Boson number in the small-$\varphi$ limit}
For the imperfect echo sequence denoted by $\hat{\mathcal{U}}\hat{\mathcal{V}}e^{-i\varphi\hat{P}}\hat{\mathcal{U}}$, the boson number of the final state is 
\begin{equation}
    \langle\hat P\rangle(\varphi) = \langle 0| e^{i\hat HT_0}e^{i\varphi\hat{P}}e^{i\hat H^{\rm e}T_1}e^{i\hat HT_0}
    \hat{P} e^{-i\hat HT_0}e^{-i\hat H^{\rm e}T_1}e^{-i\varphi\hat{P}}e^{-i\hat HT_0}|0\rangle.
\end{equation}
Using the decomposition $e^{-i\hat HT_0}=\hat{\mathcal{S}}_1\hat \Lambda \hat{\mathcal{S}}_2$ and $e^{-i\hat H^{\rm e}T_1} = \hat{\mathcal{S}}_2^{-1}\hat \Omega\hat{\mathcal{S}}_1^{-1}$, where $\hat{\mathcal{S}}_1,\hat{\mathcal{S}}_2$ are the unitary parts, $\hat \Lambda$ is the squeezing part, and $\hat \Omega=e^{-i\pi\hat{P}/2}$, to leading nonvanishing order of $\varphi$, we find
\begin{equation}
   \begin{split}
    \langle\hat P\rangle(\varphi) = & \varphi^2\langle 0| \hat \Lambda \hat{P} \hat \Lambda^{-1}\hat{P} \hat \Lambda \hat{P}  \hat \Lambda^{-1}|0\rangle+O(\varphi^3).\\
   \end{split}
\end{equation}
The identity $\hat \Lambda\hat \Omega \hat \Lambda = \hat \Omega$ is used.
Notice 
\begin{equation}
    \hat \Lambda \hat{P} \hat \Lambda^{-1} = \frac{1}{2}\sum_{j=1}^{2N}\hat{\xi}_j^2 e^{2\Lambda_j}-\frac{N}{2},
\end{equation}
where the $\Lambda_i$ are given by Eq.~(7) of the main text.
We then find
\begin{equation}
   \begin{split}
   \langle\hat P\rangle(\varphi) = & \frac{\varphi^2}{8} \bigg(\sum_{i,j,k}e^{2\Lambda_i+2\Lambda_k}\langle 0| \hat{\xi}_i^2\hat{\xi}_j^2\hat{\xi}_k^2|0\rangle-N\sum_{i,k}e^{2\Lambda_i+2\Lambda_k}\langle 0| \hat{\xi}_i^2\hat{\xi}_k^2|0\rangle\bigg).\\
   \end{split}
\end{equation}
Using Wick's theorem, $\wick{\c2{\hat\xi^m}\c2{\hat\xi^n}}=\frac{1}{2}(\delta^{mn}+i\Omega^{mn})$, the expressions above can be evaluated by contractions.
The contractions in the first term fall into three classes: \((ii)(jj)(kk)\), with one configuration; \((ij)(ij)(kk)\), with six configurations; and \((ij)(jk)(ik)\), with eight configurations. 
The results of contractions read
\begin{equation}
   \sum_{i,j,k}e^{2\Lambda_i+2\Lambda_k}\langle 0| \hat{\xi}_i^2\hat{\xi}_j^2\hat{\xi}_k^2|0\rangle = {\frac{1}{4} \bigg(N\sum_{i,k}e^{2\Lambda_i+2\Lambda_k}+2N\sum_{i}(e^{4\Lambda_i}-1)+8\sum_{i}e^{4\Lambda_i}-16N\bigg)}.
\end{equation}
The second term is 
\begin{equation}
     -N\sum_{i,k}e^{2\Lambda_i+2\Lambda_k}\langle 0| \hat{\xi}_i^2\hat{\xi}_k^2|0\rangle =-\frac{1}{4}\bigg(N\sum_{i,k}e^{2\Lambda_i+2\Lambda_k}+2N\sum_{i}(e^{4\Lambda_i}-1)\bigg).
\end{equation}
The final result is 
\begin{equation}
   \begin{split}
        \langle\hat P\rangle(\varphi) = \frac{\varphi^2}{4} \sum_{i=1}^{2N}\bigg(e^{4\Lambda_i}-1\bigg)=\frac{\varphi^2}{2}F_Q. \\
   \end{split}
\end{equation}
We have used Eqs.~(4,7) in the main text to obtain the expression of $F_Q$. 

\section{Fisher information of the Sp$(2N,R)$ interferometer}
For the echo sequence denoted by $UVU$, the final probe state is 
\begin{equation}
   \ket{\psi(\varphi)}=e^{-i\hat HT_0}e^{-i\hat H^{\rm e}T_1}e^{-i\varphi\hat{P}}e^{-i\hat HT_0}\ket{0}.
\end{equation}
Its quantum Fisher information is the same as that of the probe state $\ket{\Psi(\varphi)}=e^{-i\varphi\hat{P}}e^{-i\hat HT_0}\ket{0}$, because the two probe states are related by a unitary transformation independent of $\varphi$.

Since $\ket{\psi(\varphi=0)}=\ket{0}$ according to the definition of the Sp$(2N,R)$ echo, projective measurement of $\hat{P}$ yields a probability distribution whose classical Fisher information (CFI) $F$ saturates the bound $F=F_Q$ for vanishing $\varphi$~\cite{Smerzi2014S}.
To show this, we consider the positive operator valued measure $\hat\Pi_0=\ket{0}\bra{0}, \hat\Pi_1=1-\ket{0}\bra{0}$, which measure the probabilities of finding the vacuum and a nonzero total number of bosons at the output port, respectively.
The CFI is 
\begin{equation}
   F(\varphi) = \sum_{k=0,1}\frac{(\partial_\varphi P_k(\varphi))^2}{P_k(\varphi)}=\sum_{k=0,1}\frac{(\partial_\varphi\bra{\psi(\varphi)}\hat\Pi_k\ket{\psi(\varphi)})^2}{\bra{\psi(\varphi)}\hat\Pi_k\ket{\psi(\varphi)}}.
\end{equation} 
We now consider the limit $F(\varphi=0)$. The $k=0$ term vanishes, and the $k=1$ term can be evaluated using de l'H\^opital's rule~\cite{Smerzi2014S},
\begin{equation}
   \lim_{\varphi\to 0} F(\varphi) = \frac{\partial_\varphi(\partial_\varphi\bra{\psi(\varphi)}\hat\Pi_1\ket{\psi(\varphi)})^2}{\partial_\varphi\bra{\psi(\varphi)}\hat\Pi_1\ket{\psi(\varphi)}}\bigg|_{\varphi=0} = 4\bra{\partial_\varphi\psi(\varphi)}\hat\Pi_1\ket{\partial_\varphi\psi(\varphi)}|_{\varphi=0}=F_Q(\varphi=0).
\end{equation}
Therefore, the Sp$(2N,R)$ interferometer provides an optimal measurement that saturates the bound $F=F_Q$ in the limit $\varphi \to 0$. 
We can also perform single mode projective measurements, where $\hat\Pi_0=\hat{\mathbb{I}}_1\otimes\dots\otimes \hat{\mathbb{I}}_{i-1}\otimes \ket{0}_i\bra{0}_i\otimes \hat{\mathbb{I}}_{i+1}\otimes\dots\otimes\hat{\mathbb{I}}_{N}$, $\hat\Pi_1 = 1-\hat\Pi_0$. 
A similar calculation shows that its CFI in the limit $\varphi \to 0$ amounts to the QFI associated with $\hat P_i=\hat a_i^\dag\hat a_i$,
\begin{equation}
     \lim_{\varphi\to 0} F_i(\varphi)  = 4[\bra{\partial_\varphi\Psi_i(\varphi)}\ket{\partial_\varphi\Psi_i(\varphi)}-|\bra{\Psi_i(\varphi)}\ket{\partial_\varphi\Psi_i(\varphi)}|^2]|_{\varphi=0},
\end{equation}
where $\ket{\Psi_i(\varphi)}=e^{-i\varphi\hat{P}_i}e^{-i\hat HT_0}\ket{0}$.

\section{ISp$(2N,R)$ echoes}

Extending the Sp$(2N,R)$ echo sequence to ISp$(2N,R)$ can be achieved by applying the full Sp$(2N,R)$ echo sequence. 
Notice that an ISp$(2N,R)$ element can be written as 
\begin{equation}
    U' = \begin{pmatrix}U & f\\0 & 1\end{pmatrix} = \begin{pmatrix}{\bf I}_{2N} & f\\0 & 1\end{pmatrix} \begin{pmatrix}U & 0\\0 & 1\end{pmatrix}.
\end{equation}
Direct calculation shows that 
\begin{equation}
   (U'V')^2\equiv\bigg[\begin{pmatrix}U & f\\0 & 1\end{pmatrix} \begin{pmatrix}V & 0\\0 & 1\end{pmatrix}\bigg]^2=\begin{pmatrix}-{\bf I}_{2N} & (UV+{\bf I}_{2N})f\\0 & 1\end{pmatrix},
\end{equation}
which creates displacements to the quadratures.
Then 
\begin{equation}
   (U'V'U'V')^2=\begin{pmatrix}-{\bf I}_{2N} & (UV+{\bf I}_{2N})f\\0 & 1\end{pmatrix}^2=  {\bf I}_{2N+1}.
\end{equation}
Unlike the Sp$(2N,R)$ interferometer, three instances of a passive process, denoted by $V'$, need to be inserted to realize an ISp$(2N,R)$ interferometer, as $U'V'U'V'U'V'U'=(V')^{-1}$ is passive and therefore leaves the vacuum invariant up to a global phase. 


\begin{thebibliography}{70}%
\linespread{1.15}\selectfont
\makeatletter
\providecommand \@ifxundefined [1]{%
 \@ifx{#1\undefined}
}%
\providecommand \@ifnum [1]{%
 \ifnum #1\expandafter \@firstoftwo
 \else \expandafter \@secondoftwo
 \fi
}%
\providecommand \@ifx [1]{%
 \ifx #1\expandafter \@firstoftwo
 \else \expandafter \@secondoftwo
 \fi
}%
\providecommand \natexlab [1]{#1}%
\providecommand \enquote  [1]{#1}%
\providecommand \bibnamefont  [1]{#1}%
\providecommand \bibfnamefont [1]{#1}%
\providecommand \citenamefont [1]{#1}%
\providecommand \href@noop [0]{\@secondoftwo}%
\providecommand \href [0]{\begingroup \@sanitize@url \@href}%
\providecommand \@href[1]{\@@startlink{#1}\@@href}%
\providecommand \@@href[1]{\endgroup#1\@@endlink}%
\providecommand \@sanitize@url [0]{\catcode `\\12\catcode `\$12\catcode
  `\&12\catcode `\#12\catcode `\^12\catcode `\_12\catcode `\%12\relax}%
\providecommand \@@startlink[1]{}%
\providecommand \@@endlink[0]{}%
\providecommand \url  [0]{\begingroup\@sanitize@url \@url }%
\providecommand \@url [1]{\endgroup\@href {#1}{\urlprefix }}%
\providecommand \urlprefix  [0]{URL }%
\providecommand \Eprint [0]{\href }%
\providecommand \doibase [0]{https://dx.doi.org}%
\providecommand \selectlanguage [0]{\@gobble}%
\providecommand \bibinfo  [0]{\@secondoftwo}%
\providecommand \bibfield  [0]{\@secondoftwo}%
\providecommand \translation [1]{[#1]}%
\providecommand \BibitemOpen [0]{}%
\providecommand \bibitemStop [0]{}%
\providecommand \bibitemNoStop [0]{.\EOS\space}%
\providecommand \EOS [0]{\spacefactor3000\relax}%
\providecommand \BibitemShut  [1]{\csname bibitem#1\endcsname}%
\let\auto@bib@innerbib\@empty
%</preamble>
\bibitem [{\citenamefont {Giovannetti}\ \emph {et~al.}(2004)\citenamefont
  {Giovannetti}, \citenamefont {Lloyd},\ and\ \citenamefont
  {Maccone}}]{Giovannetti2004}%
  \BibitemOpen
  \bibfield  {author} {\bibinfo {author} {\bibfnamefont {V.}~\bibnamefont
  {Giovannetti}}, \bibinfo {author} {\bibfnamefont {S.}~\bibnamefont {Lloyd}},
  \ and\ \bibinfo {author} {\bibfnamefont {L.}~\bibnamefont {Maccone}},\
  }\bibfield  {title} {\bibinfo {title} {Quantum-enhanced measurements: Beating
  the standard quantum limit},\ }\href {\doibase/10.1126/science.1104149}
  {\bibfield  {journal} {\bibinfo  {journal} {Science}\ }\textbf {\bibinfo
  {volume} {306}},\ \bibinfo {pages} {1330} (\bibinfo {year}
  {2004})}\BibitemShut {NoStop}%
\bibitem [{\citenamefont {Giovannetti}\ \emph {et~al.}(2011)\citenamefont
  {Giovannetti}, \citenamefont {Lloyd},\ and\ \citenamefont
  {Maccone}}]{Giovannetti2011}%
  \BibitemOpen
  \bibfield  {author} {\bibinfo {author} {\bibfnamefont {V.}~\bibnamefont
  {Giovannetti}}, \bibinfo {author} {\bibfnamefont {S.}~\bibnamefont {Lloyd}},
  \ and\ \bibinfo {author} {\bibfnamefont {L.}~\bibnamefont {Maccone}},\
  }\bibfield  {title} {\bibinfo {title} {Advances in quantum metrology},\
  }\href {\doibase/10.1038/nphoton.2011.35} {\bibfield  {journal} {\bibinfo
  {journal} {Nature Photonics}\ }\textbf {\bibinfo {volume} {5}},\ \bibinfo
  {pages} {222} (\bibinfo {year} {2011})}\BibitemShut {NoStop}%
\bibitem [{\citenamefont {Degen}\ \emph {et~al.}(2017)\citenamefont {Degen},
  \citenamefont {Reinhard},\ and\ \citenamefont {Cappellaro}}]{Degen2017}%
  \BibitemOpen
  \bibfield  {author} {\bibinfo {author} {\bibfnamefont {C.~L.}\ \bibnamefont
  {Degen}}, \bibinfo {author} {\bibfnamefont {F.}~\bibnamefont {Reinhard}}, \
  and\ \bibinfo {author} {\bibfnamefont {P.}~\bibnamefont {Cappellaro}},\
  }\bibfield  {title} {\bibinfo {title} {Quantum sensing},\ }\href
  {\doibase/10.1103/RevModPhys.89.035002} {\bibfield  {journal} {\bibinfo
  {journal} {Reviews of Modern Physics}\ }\textbf {\bibinfo {volume} {89}},\ \bibinfo
  {pages} {035002} (\bibinfo {year} {2017})}\BibitemShut {NoStop}%
\bibitem [{\citenamefont {Pezzè}\ \emph {et~al.}(2018)\citenamefont {Pezzè},
  \citenamefont {Smerzi}, \citenamefont {Oberthaler}, \citenamefont {Schmied},\
  and\ \citenamefont {Treutlein}}]{Pezz2018}%
  \BibitemOpen
  \bibfield  {author} {\bibinfo {author} {\bibfnamefont {L.}~\bibnamefont
  {Pezzè}}, \bibinfo {author} {\bibfnamefont {A.}~\bibnamefont {Smerzi}},
  \bibinfo {author} {\bibfnamefont {M.~K.}\ \bibnamefont {Oberthaler}},
  \bibinfo {author} {\bibfnamefont {R.}~\bibnamefont {Schmied}}, \ and\
  \bibinfo {author} {\bibfnamefont {P.}~\bibnamefont {Treutlein}},\ }\bibfield
  {title} {\bibinfo {title} {Quantum metrology with nonclassical states of
  atomic ensembles},\ }\href {\doibase/10.1103/revmodphys.90.035005} {\bibfield
   {journal} {\bibinfo  {journal} {Reviews of Modern Physics}\ }\textbf
  {\bibinfo {volume} {90}},\ \bibinfo {pages} {035005} (\bibinfo {year} {2018})}\BibitemShut {NoStop}%
\bibitem [{\citenamefont {Helstrom}(1969)}]{Helstrom1969}%
  \BibitemOpen
  \bibfield  {author} {\bibinfo {author} {\bibfnamefont {C.~W.}\ \bibnamefont
  {Helstrom}},\ }\bibfield  {title} {\bibinfo {title} {Quantum detection and
  estimation theory},\ }\href {\doibase/10.1007/bf01007479} {\bibfield
  {journal} {\bibinfo  {journal} {Journal of Statistical Physics}\ }\textbf
  {\bibinfo {volume} {1}},\ \bibinfo {pages} {231} (\bibinfo {year}
  {1969})}\BibitemShut {NoStop}%
\bibitem [{\citenamefont {Caves}(1981)}]{Caves1981}%
  \BibitemOpen
  \bibfield  {author} {\bibinfo {author} {\bibfnamefont {C.~M.}\ \bibnamefont
  {Caves}},\ }\bibfield  {title} {\bibinfo {title} {Quantum-mechanical noise in
  an interferometer},\ }\href {\doibase/10.1103/physrevd.23.1693} {\bibfield
  {journal} {\bibinfo  {journal} {Physical Review D}\ }\textbf {\bibinfo
  {volume} {23}},\ \bibinfo {pages} {1693} (\bibinfo {year}
  {1981})}\BibitemShut {NoStop}%
\bibitem [{\citenamefont {Yurke}\ \emph {et~al.}(1986)\citenamefont {Yurke},
  \citenamefont {McCall},\ and\ \citenamefont {Klauder}}]{Yurke1986}%
  \BibitemOpen
  \bibfield  {author} {\bibinfo {author} {\bibfnamefont {B.}~\bibnamefont
  {Yurke}}, \bibinfo {author} {\bibfnamefont {S.~L.}\ \bibnamefont {McCall}}, \
  and\ \bibinfo {author} {\bibfnamefont {J.~R.}\ \bibnamefont {Klauder}},\
  }\bibfield  {title} {\bibinfo {title} {SU(2) and SU(1, 1) interferometers},\
  }\href {\doibase/10.1103/physreva.33.4033} {\bibfield  {journal} {\bibinfo
  {journal} {Physical Review A}\ }\textbf {\bibinfo {volume} {33}},\ \bibinfo
  {pages} {4033} (\bibinfo {year} {1986})}\BibitemShut {NoStop}%
\bibitem [{\citenamefont {Rarity}\ \emph {et~al.}(1990)\citenamefont {Rarity},
  \citenamefont {Tapster}, \citenamefont {Jakeman}, \citenamefont {Larchuk},
  \citenamefont {Campos}, \citenamefont {Teich},\ and\ \citenamefont
  {Saleh}}]{Rarity1990}%
  \BibitemOpen
  \bibfield  {author} {\bibinfo {author} {\bibfnamefont {J.}~\bibnamefont
  {Rarity}}, \bibinfo {author} {\bibfnamefont {P.}~\bibnamefont {Tapster}},
  \bibinfo {author} {\bibfnamefont {E.}~\bibnamefont {Jakeman}}, \bibinfo
  {author} {\bibfnamefont {T.}~\bibnamefont {Larchuk}}, \bibinfo {author}
  {\bibfnamefont {R.}~\bibnamefont {Campos}}, \bibinfo {author} {\bibfnamefont
  {M.}~\bibnamefont {Teich}}, \ and\ \bibinfo {author} {\bibfnamefont
  {B.}~\bibnamefont {Saleh}},\ }\bibfield  {title} {\bibinfo {title}
  {Two-photon interference in a Mach-Zehnder interferometer},\ }\href
  {\doibase/10.1103/physrevlett.65.1348} {\bibfield  {journal} {\bibinfo
  {journal} {Physical Review Letters}\ }\textbf {\bibinfo {volume} {65}},\
  \bibinfo {pages} {1348} (\bibinfo {year} {1990})}\BibitemShut {NoStop}%
\bibitem [{\citenamefont {Nagata}\ \emph {et~al.}(2007)\citenamefont {Nagata},
  \citenamefont {Okamoto}, \citenamefont {O’Brien}, \citenamefont {Sasaki},\
  and\ \citenamefont {Takeuchi}}]{Nagata2007}%
  \BibitemOpen
  \bibfield  {author} {\bibinfo {author} {\bibfnamefont {T.}~\bibnamefont
  {Nagata}}, \bibinfo {author} {\bibfnamefont {R.}~\bibnamefont {Okamoto}},
  \bibinfo {author} {\bibfnamefont {J.~L.}\ \bibnamefont {O’Brien}}, \bibinfo
  {author} {\bibfnamefont {K.}~\bibnamefont {Sasaki}}, \ and\ \bibinfo {author}
  {\bibfnamefont {S.}~\bibnamefont {Takeuchi}},\ }\bibfield  {title} {\bibinfo
  {title} {Beating the standard quantum limit with four-entangled photons},\
  }\href {\doibase/10.1126/science.1138007} {\bibfield  {journal} {\bibinfo
  {journal} {Science}\ }\textbf {\bibinfo {volume} {316}},\ \bibinfo {pages}
  {726} (\bibinfo {year} {2007})}\BibitemShut {NoStop}%
\bibitem [{\citenamefont {Aasi~et al}(2013)}]{Aasi2013}%
  \BibitemOpen
  \bibfield  {author} {\bibinfo {author} {\bibfnamefont {J.}~\bibnamefont
  {Aasi}},\ {\bibfnamefont {J.}~\bibnamefont{Abadie}},\ {\bibfnamefont {B.}~\bibnamefont{Abbott~et al.}},\ }\bibfield  {title} {\bibinfo {title} {Enhanced sensitivity of the LIGO gravitational wave detector by using squeezed states of light},\
  }\href {\doibase/10.1038/nphoton.2013.177} {\bibfield  {journal} {\bibinfo
  {journal} {Nature Photonics}\ }\textbf {\bibinfo {volume} {7}},\ \bibinfo
  {pages} {613} (\bibinfo {year} {2013})}\BibitemShut {NoStop}%
\bibitem [{\citenamefont {Kitagawa}\ and\ \citenamefont
  {Ueda}(1993)}]{Kitagawa1992}%
  \BibitemOpen
  \bibfield  {author} {\bibinfo {author} {\bibfnamefont {M.}~\bibnamefont
  {Kitagawa}}\ and\ \bibinfo {author} {\bibfnamefont {M.}~\bibnamefont
  {Ueda}},\ }\bibfield  {title} {\bibinfo {title} {Squeezed spin states},\
  }\href {\doibase/10.1103/physreva.47.5138} {\bibfield  {journal} {\bibinfo
  {journal} {Physical Review A}\ }\textbf {\bibinfo {volume} {47}},\ \bibinfo
  {pages} {5138} (\bibinfo {year} {1993})}\BibitemShut {NoStop}%
\bibitem [{\citenamefont {Wineland}\ \emph {et~al.}(1992)\citenamefont
  {Wineland}, \citenamefont {Bollinger}, \citenamefont {Itano}, \citenamefont
  {Moore},\ and\ \citenamefont {Heinzen}}]{Wineland1992}%
  \BibitemOpen
  \bibfield  {author} {\bibinfo {author} {\bibfnamefont {D.~J.}\ \bibnamefont
  {Wineland}}, \bibinfo {author} {\bibfnamefont {J.~J.}\ \bibnamefont
  {Bollinger}}, \bibinfo {author} {\bibfnamefont {W.~M.}\ \bibnamefont
  {Itano}}, \bibinfo {author} {\bibfnamefont {F.~L.}\ \bibnamefont {Moore}}, \
  and\ \bibinfo {author} {\bibfnamefont {D.~J.}\ \bibnamefont {Heinzen}},\
  }\bibfield  {title} {\bibinfo {title} {Spin squeezing and reduced quantum
  noise in spectroscopy},\ }\href {\doibase/10.1103/physreva.46.r6797}
  {\bibfield  {journal} {\bibinfo  {journal} {Physical Review A}\ }\textbf
  {\bibinfo {volume} {46}},\ \bibinfo {pages} {R6797} (\bibinfo {year}
  {1992})}\BibitemShut {NoStop}%
\bibitem [{\citenamefont {Estève}\ \emph {et~al.}(2008)\citenamefont
  {Estève}, \citenamefont {Gross}, \citenamefont {Weller}, \citenamefont
  {Giovanazzi},\ and\ \citenamefont {Oberthaler}}]{Estve2008}%
  \BibitemOpen
  \bibfield  {author} {\bibinfo {author} {\bibfnamefont {J.}~\bibnamefont
  {Estève}}, \bibinfo {author} {\bibfnamefont {C.}~\bibnamefont {Gross}},
  \bibinfo {author} {\bibfnamefont {A.}~\bibnamefont {Weller}}, \bibinfo
  {author} {\bibfnamefont {S.}~\bibnamefont {Giovanazzi}}, \ and\ \bibinfo
  {author} {\bibfnamefont {M.~K.}\ \bibnamefont {Oberthaler}},\ }\bibfield
  {title} {\bibinfo {title} {Squeezing and entanglement in a Bose-Einstein
  condensate},\ }\href {\doibase/10.1038/nature07332} {\bibfield  {journal}
  {\bibinfo  {journal} {Nature}\ }\textbf {\bibinfo {volume} {455}},\ \bibinfo
  {pages} {1216} (\bibinfo {year} {2008})}\BibitemShut {NoStop}%
\bibitem [{\citenamefont {Mao}\ \emph {et~al.}(2023)\citenamefont {Mao},
  \citenamefont {Liu}, \citenamefont {Li}, \citenamefont {Cao}, \citenamefont
  {Chen}, \citenamefont {Xu}, \citenamefont {Tey}, \citenamefont {Huang},\ and\
  \citenamefont {You}}]{Mao2023}%
  \BibitemOpen
  \bibfield  {author} {\bibinfo {author} {\bibfnamefont {T.-W.}\ \bibnamefont
  {Mao}}, \bibinfo {author} {\bibfnamefont {Q.}~\bibnamefont {Liu}}, \bibinfo
  {author} {\bibfnamefont {X.-W.}\ \bibnamefont {Li}}, \bibinfo {author}
  {\bibfnamefont {J.-H.}\ \bibnamefont {Cao}}, \bibinfo {author} {\bibfnamefont
  {F.}~\bibnamefont {Chen}}, \bibinfo {author} {\bibfnamefont {W.-X.}\
  \bibnamefont {Xu}}, \bibinfo {author} {\bibfnamefont {M.~K.}\ \bibnamefont
  {Tey}}, \bibinfo {author} {\bibfnamefont {Y.-X.}\ \bibnamefont {Huang}}, \
  and\ \bibinfo {author} {\bibfnamefont {L.}~\bibnamefont {You}},\ }\bibfield
  {title} {\bibinfo {title} {Quantum-enhanced sensing by echoing spin-nematic
  squeezing in atomic Bose-Einstein condensate},\ }\href
  {\doibase/10.1038/s41567-023-02168-3} {\bibfield  {journal} {\bibinfo
  {journal} {Nature Physics}\ }\textbf {\bibinfo {volume} {19}},\ \bibinfo
  {pages} {1585} (\bibinfo {year} {2023})}\BibitemShut {NoStop}%
\bibitem [{\citenamefont {Franke}\ \emph {et~al.}(2023)\citenamefont {Franke},
  \citenamefont {Muleady}, \citenamefont {Kaubruegger}, \citenamefont {Kranzl},
  \citenamefont {Blatt}, \citenamefont {Rey}, \citenamefont {Joshi},\ and\
  \citenamefont {Roos}}]{Franke2023}%
  \BibitemOpen
  \bibfield  {author} {\bibinfo {author} {\bibfnamefont {J.}~\bibnamefont
  {Franke}}, \bibinfo {author} {\bibfnamefont {S.~R.}\ \bibnamefont {Muleady}},
  \bibinfo {author} {\bibfnamefont {R.}~\bibnamefont {Kaubruegger}}, \bibinfo
  {author} {\bibfnamefont {F.}~\bibnamefont {Kranzl}}, \bibinfo {author}
  {\bibfnamefont {R.}~\bibnamefont {Blatt}}, \bibinfo {author} {\bibfnamefont
  {A.~M.}\ \bibnamefont {Rey}}, \bibinfo {author} {\bibfnamefont {M.~K.}\
  \bibnamefont {Joshi}}, \ and\ \bibinfo {author} {\bibfnamefont {C.~F.}\
  \bibnamefont {Roos}},\ }\bibfield  {title} {\bibinfo {title}
  {Quantum-enhanced sensing on optical transitions through finite-range
  interactions},\ }\href {\doibase/10.1038/s41586-023-06472-z} {\bibfield
  {journal} {\bibinfo  {journal} {Nature}\ }\textbf {\bibinfo {volume} {621}},\
  \bibinfo {pages} {740} (\bibinfo {year} {2023})}\BibitemShut {NoStop}%
\bibitem [{\citenamefont {Eckner}\ \emph {et~al.}(2023)\citenamefont {Eckner},
  \citenamefont {Darkwah~Oppong}, \citenamefont {Cao}, \citenamefont {Young},
  \citenamefont {Milner}, \citenamefont {Robinson}, \citenamefont {Ye},\ and\
  \citenamefont {Kaufman}}]{Eckner2023}%
  \BibitemOpen
  \bibfield  {author} {\bibinfo {author} {\bibfnamefont {W.~J.}\ \bibnamefont
  {Eckner}}, \bibinfo {author} {\bibfnamefont {N.}~\bibnamefont
  {Darkwah~Oppong}}, \bibinfo {author} {\bibfnamefont {A.}~\bibnamefont {Cao}},
  \bibinfo {author} {\bibfnamefont {A.~W.}\ \bibnamefont {Young}}, \bibinfo
  {author} {\bibfnamefont {W.~R.}\ \bibnamefont {Milner}}, \bibinfo {author}
  {\bibfnamefont {J.~M.}\ \bibnamefont {Robinson}}, \bibinfo {author}
  {\bibfnamefont {J.}~\bibnamefont {Ye}}, \ and\ \bibinfo {author}
  {\bibfnamefont {A.~M.}\ \bibnamefont {Kaufman}},\ }\bibfield  {title}
  {\bibinfo {title} {Realizing spin squeezing with Rydberg interactions in an
  optical clock},\ }\href {\doibase/10.1038/s41586-023-06360-6} {\bibfield
  {journal} {\bibinfo  {journal} {Nature}\ }\textbf {\bibinfo {volume} {621}},\
  \bibinfo {pages} {734} (\bibinfo {year} {2023})}\BibitemShut {NoStop}%
\bibitem [{\citenamefont {Schumaker}(1986)}]{Schumaker1986}%
  \BibitemOpen
  \bibfield  {author} {\bibinfo {author} {\bibfnamefont {B.~L.}\ \bibnamefont
  {Schumaker}},\ }\bibfield  {title} {\bibinfo {title} {Quantum mechanical pure
  states with Gaussian wave functions},\ }\href
  {\doibase/10.1016/0370-1573(86)90179-1} {\bibfield  {journal} {\bibinfo
  {journal} {Physics Reports}\ }\textbf {\bibinfo {volume} {135}},\ \bibinfo
  {pages} {317} (\bibinfo {year} {1986})}\BibitemShut {NoStop}%
\bibitem [{\citenamefont {Braunstein}\ and\ \citenamefont {van
  Loock}(2005)}]{Braunstein2005_Review}%
  \BibitemOpen
  \bibfield  {author} {\bibinfo {author} {\bibfnamefont {S.~L.}\ \bibnamefont
  {Braunstein}}\ and\ \bibinfo {author} {\bibfnamefont {P.}~\bibnamefont {van
  Loock}},\ }\bibfield  {title} {\bibinfo {title} {Quantum information with
  continuous variables},\ }\href {\doibase/10.1103/revmodphys.77.513}
  {\bibfield  {journal} {\bibinfo  {journal} {Reviews of Modern Physics}\
  }\textbf {\bibinfo {volume} {77}},\ \bibinfo {pages} {513} (\bibinfo {year}
  {2005})}\BibitemShut {NoStop}%
\bibitem [{\citenamefont {Monras}(2006)}]{Monras2006}%
  \BibitemOpen
  \bibfield  {author} {\bibinfo {author} {\bibfnamefont {A.}~\bibnamefont
  {Monras}},\ }\bibfield  {title} {\bibinfo {title} {Optimal phase measurements
  with pure Gaussian states},\ }\href {\doibase/10.1103/physreva.73.033821}
  {\bibfield  {journal} {\bibinfo  {journal} {Physical Review A}\ }\textbf
  {\bibinfo {volume} {73}},\ \bibinfo {pages} {033821} (\bibinfo {year} {2006})}\BibitemShut {NoStop}%
\bibitem [{\citenamefont {Pinel}\ \emph {et~al.}(2012)\citenamefont {Pinel},
  \citenamefont {Fade}, \citenamefont {Braun}, \citenamefont {Jian},
  \citenamefont {Treps},\ and\ \citenamefont {Fabre}}]{Pinel2012}%
  \BibitemOpen
  \bibfield  {author} {\bibinfo {author} {\bibfnamefont {O.}~\bibnamefont
  {Pinel}}, \bibinfo {author} {\bibfnamefont {J.}~\bibnamefont {Fade}},
  \bibinfo {author} {\bibfnamefont {D.}~\bibnamefont {Braun}}, \bibinfo
  {author} {\bibfnamefont {P.}~\bibnamefont {Jian}}, \bibinfo {author}
  {\bibfnamefont {N.}~\bibnamefont {Treps}}, \ and\ \bibinfo {author}
  {\bibfnamefont {C.}~\bibnamefont {Fabre}},\ }\bibfield  {title} {\bibinfo
  {title} {Ultimate sensitivity of precision measurements with intense Gaussian
  quantum light: A multimodal approach},\ }\href
  {\doibase/10.1103/physreva.85.010101} {\bibfield  {journal} {\bibinfo
  {journal} {Physical Review A}\ }\textbf {\bibinfo {volume} {85}},\ \bibinfo {pages} {010101} (\bibinfo
  {year} {2012})}\BibitemShut {NoStop}%
\bibitem [{\citenamefont {Pinel}\ \emph {et~al.}(2013)\citenamefont {Pinel},
  \citenamefont {Jian}, \citenamefont {Treps}, \citenamefont {Fabre},\ and\
  \citenamefont {Braun}}]{Pinel2013}%
  \BibitemOpen
  \bibfield  {author} {\bibinfo {author} {\bibfnamefont {O.}~\bibnamefont
  {Pinel}}, \bibinfo {author} {\bibfnamefont {P.}~\bibnamefont {Jian}},
  \bibinfo {author} {\bibfnamefont {N.}~\bibnamefont {Treps}}, \bibinfo
  {author} {\bibfnamefont {C.}~\bibnamefont {Fabre}}, \ and\ \bibinfo {author}
  {\bibfnamefont {D.}~\bibnamefont {Braun}},\ }\bibfield  {title} {\bibinfo
  {title} {Quantum parameter estimation using general single-mode Gaussian
  states},\ }\href {\doibase/10.1103/physreva.88.040102} {\bibfield  {journal}
  {\bibinfo  {journal} {Physical Review A}\ }\textbf {\bibinfo {volume} {88}},\ \bibinfo {pages} {040102}
  (\bibinfo {year} {2013})}\BibitemShut {NoStop}%
\bibitem [{\citenamefont {Weedbrook}\ \emph {et~al.}(2012)\citenamefont
  {Weedbrook}, \citenamefont {Pirandola}, \citenamefont {García-Patrón},
  \citenamefont {Cerf}, \citenamefont {Ralph}, \citenamefont {Shapiro},\ and\
  \citenamefont {Lloyd}}]{Weedbrook2012}%
  \BibitemOpen
  \bibfield  {author} {\bibinfo {author} {\bibfnamefont {C.}~\bibnamefont
  {Weedbrook}}, \bibinfo {author} {\bibfnamefont {S.}~\bibnamefont
  {Pirandola}}, \bibinfo {author} {\bibfnamefont {R.}~\bibnamefont
  {García-Patrón}}, \bibinfo {author} {\bibfnamefont {N.~J.}\ \bibnamefont
  {Cerf}}, \bibinfo {author} {\bibfnamefont {T.~C.}\ \bibnamefont {Ralph}},
  \bibinfo {author} {\bibfnamefont {J.~H.}\ \bibnamefont {Shapiro}}, \ and\
  \bibinfo {author} {\bibfnamefont {S.}~\bibnamefont {Lloyd}},\ }\bibfield
  {title} {\bibinfo {title} {Gaussian quantum information},\ }\href
  {\doibase/10.1103/revmodphys.84.621} {\bibfield  {journal} {\bibinfo
  {journal} {Reviews of Modern Physics}\ }\textbf {\bibinfo {volume} {84}},\
  \bibinfo {pages} {621} (\bibinfo {year} {2012})}\BibitemShut {NoStop}%
\bibitem [{\citenamefont {Albert}\ \emph {et~al.}(2016)\citenamefont {Albert},
  \citenamefont {Shu}, \citenamefont {Krastanov}, \citenamefont {Shen},
  \citenamefont {Liu}, \citenamefont {Yang}, \citenamefont {Schoelkopf},
  \citenamefont {Mirrahimi}, \citenamefont {Devoret},\ and\ \citenamefont
  {Jiang}}]{Albert2016}%
  \BibitemOpen
  \bibfield  {author} {\bibinfo {author} {\bibfnamefont {V.~V.}\ \bibnamefont
  {Albert}}, \bibinfo {author} {\bibfnamefont {C.}~\bibnamefont {Shu}},
  \bibinfo {author} {\bibfnamefont {S.}~\bibnamefont {Krastanov}}, \bibinfo
  {author} {\bibfnamefont {C.}~\bibnamefont {Shen}}, \bibinfo {author}
  {\bibfnamefont {R.-B.}\ \bibnamefont {Liu}}, \bibinfo {author} {\bibfnamefont
  {Z.-B.}\ \bibnamefont {Yang}}, \bibinfo {author} {\bibfnamefont {R.~J.}\
  \bibnamefont {Schoelkopf}}, \bibinfo {author} {\bibfnamefont
  {M.}~\bibnamefont {Mirrahimi}}, \bibinfo {author} {\bibfnamefont {M.~H.}\
  \bibnamefont {Devoret}}, \ and\ \bibinfo {author} {\bibfnamefont
  {L.}~\bibnamefont {Jiang}},\ }\bibfield  {title} {\bibinfo {title} {Holonomic
  quantum control with continuous variable systems},\ }\href
  {\doibase/10.1103/PhysRevLett.116.140502} {\bibfield  {journal} {\bibinfo
  {journal} {Physical Review Letters}\ }\textbf {\bibinfo {volume} {116}},\ \bibinfo
  {pages} {140502} (\bibinfo {year} {2016})}\BibitemShut {NoStop}%
\bibitem [{\citenamefont {Matsubara}\ \emph {et~al.}(2019)\citenamefont
  {Matsubara}, \citenamefont {Facchi}, \citenamefont {Giovannetti},\ and\
  \citenamefont {Yuasa}}]{Matsubara2019}%
  \BibitemOpen
  \bibfield  {author} {\bibinfo {author} {\bibfnamefont {T.}~\bibnamefont
  {Matsubara}}, \bibinfo {author} {\bibfnamefont {P.}~\bibnamefont {Facchi}},
  \bibinfo {author} {\bibfnamefont {V.}~\bibnamefont {Giovannetti}}, \ and\
  \bibinfo {author} {\bibfnamefont {K.}~\bibnamefont {Yuasa}},\ }\bibfield
  {title} {\bibinfo {title} {Optimal Gaussian metrology for generic multimode
  interferometric circuit},\ }\href {\doibase/10.1088/1367-2630/ab0604}
  {\bibfield  {journal} {\bibinfo  {journal} {New Journal of Physics}\ }\textbf
  {\bibinfo {volume} {21}},\ \bibinfo {pages} {033014} (\bibinfo {year}
  {2019})}\BibitemShut {NoStop}%
\bibitem [{\citenamefont {Polkovnikov}(2010)}]{Polkovnikov2010}%
  \BibitemOpen
  \bibfield  {author} {\bibinfo {author} {\bibfnamefont {A.}~\bibnamefont
  {Polkovnikov}},\ }\bibfield  {title} {\bibinfo {title} {Phase space
  representation of quantum dynamics},\ }\href
  {\doibase/10.1016/j.aop.2010.02.006} {\bibfield  {journal} {\bibinfo
  {journal} {Annals of Physics}\ }\textbf {\bibinfo {volume} {325}},\ \bibinfo
  {pages} {1790} (\bibinfo {year} {2010})}\BibitemShut {NoStop}%
\bibitem [{\citenamefont {Slim}\ \emph {et~al.}(2024)\citenamefont {Slim},
  \citenamefont {Wanjura}, \citenamefont {Brunelli}, \citenamefont {del Pino},
  \citenamefont {Nunnenkamp},\ and\ \citenamefont {Verhagen}}]{Slim2024}%
  \BibitemOpen
  \bibfield  {author} {\bibinfo {author} {\bibfnamefont {J.~J.}\ \bibnamefont
  {Slim}}, \bibinfo {author} {\bibfnamefont {C.~C.}\ \bibnamefont {Wanjura}},
  \bibinfo {author} {\bibfnamefont {M.}~\bibnamefont {Brunelli}}, \bibinfo
  {author} {\bibfnamefont {J.}~\bibnamefont {del Pino}}, \bibinfo {author}
  {\bibfnamefont {A.}~\bibnamefont {Nunnenkamp}}, \ and\ \bibinfo {author}
  {\bibfnamefont {E.}~\bibnamefont {Verhagen}},\ }\bibfield  {title} {\bibinfo
  {title} {Optomechanical realization of the bosonic Kitaev chain},\ }\href
  {\doibase/10.1038/s41586-024-07174-w} {\bibfield  {journal} {\bibinfo
  {journal} {Nature}\ }\textbf {\bibinfo {volume} {627}},\ \bibinfo {pages}
  {767} (\bibinfo {year} {2024})}\BibitemShut {NoStop}%
\bibitem [{\citenamefont {Busnaina}\ \emph {et~al.}(2024)\citenamefont
  {Busnaina}, \citenamefont {Shi}, \citenamefont {McDonald}, \citenamefont
  {Dubyna}, \citenamefont {Nsanzineza}, \citenamefont {Hung}, \citenamefont
  {Chang}, \citenamefont {Clerk},\ and\ \citenamefont {Wilson}}]{Clerk2024}%
  \BibitemOpen
  \bibfield  {author} {\bibinfo {author} {\bibfnamefont {J.~H.}\ \bibnamefont
  {Busnaina}}, \bibinfo {author} {\bibfnamefont {Z.}~\bibnamefont {Shi}},
  \bibinfo {author} {\bibfnamefont {A.}~\bibnamefont {McDonald}}, \bibinfo
  {author} {\bibfnamefont {D.}~\bibnamefont {Dubyna}}, \bibinfo {author}
  {\bibfnamefont {I.}~\bibnamefont {Nsanzineza}}, \bibinfo {author}
  {\bibfnamefont {J.~S.~C.}\ \bibnamefont {Hung}}, \bibinfo {author}
  {\bibfnamefont {C.~W.~S.}\ \bibnamefont {Chang}}, \bibinfo {author}
  {\bibfnamefont {A.~A.}\ \bibnamefont {Clerk}}, \ and\ \bibinfo {author}
  {\bibfnamefont {C.~M.}\ \bibnamefont {Wilson}},\ }\bibfield  {title}
  {\bibinfo {title} {Quantum simulation of the bosonic Kitaev chain},\ }\href
  {\doibase/10.1038/s41467-024-47186-8} {\bibfield  {journal} {\bibinfo
  {journal} {Nature Communications}\ }\textbf {\bibinfo {volume} {15}},\ \bibinfo {pages}{3065} (\bibinfo {year} {2024})}\BibitemShut {NoStop}%
\bibitem [{\citenamefont {Szczykulska}\ \emph {et~al.}(2016)\citenamefont
  {Szczykulska}, \citenamefont {Baumgratz},\ and\ \citenamefont
  {Datta}}]{Szczykulska2016}%
  \BibitemOpen
  \bibfield  {author} {\bibinfo {author} {\bibfnamefont {M.}~\bibnamefont
  {Szczykulska}}, \bibinfo {author} {\bibfnamefont {T.}~\bibnamefont
  {Baumgratz}}, \ and\ \bibinfo {author} {\bibfnamefont {A.}~\bibnamefont
  {Datta}},\ }\bibfield  {title} {\bibinfo {title} {Multi-parameter quantum
  metrology},\ }\href {\doibase/10.1080/23746149.2016.1230476} {\bibfield
  {journal} {\bibinfo  {journal} {Advances in Physics: X}\ }\textbf {\bibinfo
  {volume} {1}},\ \bibinfo {pages} {621-639} (\bibinfo {year}
  {2016})}\BibitemShut {NoStop}%
\bibitem [{\citenamefont {Proctor}\ \emph {et~al.}(2018)\citenamefont
  {Proctor}, \citenamefont {Knott},\ and\ \citenamefont
  {Dunningham}}]{Jacob2018}%
  \BibitemOpen
  \bibfield  {author} {\bibinfo {author} {\bibfnamefont {T.~J.}\ \bibnamefont
  {Proctor}}, \bibinfo {author} {\bibfnamefont {P.~A.}\ \bibnamefont {Knott}},
  \ and\ \bibinfo {author} {\bibfnamefont {J.~A.}\ \bibnamefont {Dunningham}},\
  }\bibfield  {title} {\bibinfo {title} {Multiparameter estimation in networked
  quantum sensors},\ }\href {\doibase/10.1103/PhysRevLett.120.080501}
  {\bibfield  {journal} {\bibinfo  {journal} {Physical Review Letters}\ }\textbf
  {\bibinfo {volume} {120}},\ \bibinfo {pages} {080501} (\bibinfo {year}
  {2018})}\BibitemShut {NoStop}%
\bibitem [{\citenamefont {Šafránek}(2018)}]{Afrnek2018}%
  \BibitemOpen
  \bibfield  {author} {\bibinfo {author} {\bibfnamefont {D.}~\bibnamefont {Šafránek}},\ }\bibfield  {title} {\bibinfo {title} {Estimation of Gaussian quantum states},\ }\href {\doibase/10.1088/1751-8121/aaf068} {\bibfield  {journal} {\bibinfo  {journal} {Journal of Physics A: Mathematical and Theoretical}\ }\textbf {\bibinfo {volume} {52}},\ \bibinfo {pages} {035304} (\bibinfo {year} {2018})}\BibitemShut {NoStop}%
\bibitem [{\citenamefont {Nichols}\ \emph {et~al.}(2018)\citenamefont {Nichols}, \citenamefont {Liuzzo-Scorpo}, \citenamefont {Knott},\ and\ \citenamefont {Adesso}}]{Adesso2018}%
  \BibitemOpen
  \bibfield  {author} {\bibinfo {author} {\bibfnamefont {R.}~\bibnamefont {Nichols}}, \bibinfo {author} {\bibfnamefont {P.}~\bibnamefont {Liuzzo-Scorpo}}, \bibinfo {author} {\bibfnamefont {P.~A.}\ \bibnamefont {Knott}}, \ and\ \bibinfo {author} {\bibfnamefont {G.}~\bibnamefont {Adesso}},\ }\bibfield  {title} {\bibinfo {title} {Multiparameter Gaussian quantum metrology},\ }\href {\doibase/10.1103/PhysRevA.98.012114} {\bibfield  {journal} {\bibinfo  {journal} {Phys. Rev. A}\ }\textbf {\bibinfo {volume} {98}},\ \bibinfo {pages} {012114} (\bibinfo {year} {2018})}\BibitemShut {NoStop}%
\bibitem [{\citenamefont {Liu}\ \emph {et~al.}(2019)\citenamefont {Liu},
  \citenamefont {Yuan}, \citenamefont {Lu},\ and\ \citenamefont
  {Wang}}]{Liu2019}%
  \BibitemOpen
  \bibfield  {author} {\bibinfo {author} {\bibfnamefont {J.}~\bibnamefont
  {Liu}}, \bibinfo {author} {\bibfnamefont {H.}~\bibnamefont {Yuan}}, \bibinfo
  {author} {\bibfnamefont {X.-M.}\ \bibnamefont {Lu}}, \ and\ \bibinfo {author}
  {\bibfnamefont {X.}~\bibnamefont {Wang}},\ }\bibfield  {title} {\bibinfo
  {title} {Quantum Fisher information matrix and multiparameter estimation},\
  }\href {\doibase/10.1088/1751-8121/ab5d4d} {\bibfield  {journal} {\bibinfo
  {journal} {Journal of Physics A: Mathematical and Theoretical}\ }\textbf
  {\bibinfo {volume} {53}},\ \bibinfo {pages} {023001} (\bibinfo {year}
  {2019})}\BibitemShut {NoStop}%
\bibitem [{\citenamefont {Zhang}\ and\ \citenamefont
  {Zhuang}(2021)}]{Zhang2021}%
  \BibitemOpen
  \bibfield  {author} {\bibinfo {author} {\bibfnamefont {Z.}~\bibnamefont
  {Zhang}}\ and\ \bibinfo {author} {\bibfnamefont {Q.}~\bibnamefont {Zhuang}},\
  }\bibfield  {title} {\bibinfo {title} {Distributed quantum sensing},\ }\href
  {\doibase/10.1088/2058-9565/abd4c3} {\bibfield  {journal} {\bibinfo
  {journal} {Quantum Science and Technology}\ }\textbf {\bibinfo {volume}
  {6}},\ \bibinfo {pages} {043001} (\bibinfo {year} {2021})}\BibitemShut
  {NoStop}%
\bibitem [{\citenamefont {Pezz\`e}\ and\ \citenamefont
  {Smerzi}(2025)}]{Smerzi2025}%
  \BibitemOpen
  \bibfield  {author} {\bibinfo {author} {\bibfnamefont {L.}~\bibnamefont
  {Pezz\`e}}\ and\ \bibinfo {author} {\bibfnamefont {A.}~\bibnamefont
  {Smerzi}},\ }\bibfield  {title} {\bibinfo {title} {Distributed quantum
  multiparameter estimation with optimal local measurements},\ }\href
  {\doibase/10.1103/f2jf-bg7g} {\bibfield  {journal} {\bibinfo  {journal}
  {Physical Review Letters}\ }\textbf {\bibinfo {volume} {135}},\ \bibinfo {pages}
  {260805} (\bibinfo {year} {2025})}\BibitemShut {NoStop}%
\bibitem [{\citenamefont {Hamilton}\ \emph {et~al.}(2017)\citenamefont
  {Hamilton}, \citenamefont {Kruse}, \citenamefont {Sansoni}, \citenamefont
  {Barkhofen}, \citenamefont {Silberhorn},\ and\ \citenamefont
  {Jex}}]{Jex2017}%
  \BibitemOpen
  \bibfield  {author} {\bibinfo {author} {\bibfnamefont {C.~S.}\ \bibnamefont
  {Hamilton}}, \bibinfo {author} {\bibfnamefont {R.}~\bibnamefont {Kruse}},
  \bibinfo {author} {\bibfnamefont {L.}~\bibnamefont {Sansoni}}, \bibinfo
  {author} {\bibfnamefont {S.}~\bibnamefont {Barkhofen}}, \bibinfo {author}
  {\bibfnamefont {C.}~\bibnamefont {Silberhorn}}, \ and\ \bibinfo {author}
  {\bibfnamefont {I.}~\bibnamefont {Jex}},\ }\bibfield  {title} {\bibinfo
  {title} {Gaussian boson sampling},\ }\href
  {\doibase/10.1103/PhysRevLett.119.170501} {\bibfield  {journal} {\bibinfo
  {journal} {Physical Review Letters}\ }\textbf {\bibinfo {volume} {119}},\ \bibinfo
  {pages} {170501} (\bibinfo {year} {2017})}\BibitemShut {NoStop}%
\bibitem [{\citenamefont {Wang}\ \emph {et~al.}(2017)\citenamefont {Wang},
  \citenamefont {He}, \citenamefont {Li}, \citenamefont {Su}, \citenamefont
  {Li}, \citenamefont {Huang}, \citenamefont {Ding}, \citenamefont {Chen},
  \citenamefont {Liu}, \citenamefont {Qin}, \citenamefont {Li}, \citenamefont
  {He}, \citenamefont {Schneider}, \citenamefont {Kamp}, \citenamefont {Peng},
  \citenamefont {H\"{o}fling}, \citenamefont {Lu},\ and\ \citenamefont
  {Pan}}]{Wang2017}%
  \BibitemOpen
  \bibfield  {author} {\bibinfo {author} {\bibfnamefont {H.}~\bibnamefont
  {Wang}}, \bibinfo {author} {\bibfnamefont {Y.}~\bibnamefont {He}}, \bibinfo
  {author} {\bibfnamefont {Y.-H.}\ \bibnamefont {Li}}, \bibinfo {author}
  {\bibfnamefont {Z.-E.}\ \bibnamefont {Su}}, \bibinfo {author} {\bibfnamefont
  {B.}~\bibnamefont {Li}}, \bibinfo {author} {\bibfnamefont {H.-L.}\
  \bibnamefont {Huang}}, \bibinfo {author} {\bibfnamefont {X.}~\bibnamefont
  {Ding}}, \bibinfo {author} {\bibfnamefont {M.-C.}\ \bibnamefont {Chen}},
  \bibinfo {author} {\bibfnamefont {C.}~\bibnamefont {Liu}}, \bibinfo {author}
  {\bibfnamefont {J.}~\bibnamefont {Qin}}, \bibinfo {author} {\bibfnamefont
  {J.-P.}\ \bibnamefont {Li}}, \bibinfo {author} {\bibfnamefont {Y.-M.}\
  \bibnamefont {He}}, \bibinfo {author} {\bibfnamefont {C.}~\bibnamefont
  {Schneider}}, \bibinfo {author} {\bibfnamefont {M.}~\bibnamefont {Kamp}},
  \bibinfo {author} {\bibfnamefont {C.-Z.}\ \bibnamefont {Peng}}, \bibinfo
  {author} {\bibfnamefont {S.}~\bibnamefont {H\"{o}fling}}, \bibinfo {author}
  {\bibfnamefont {C.-Y.}\ \bibnamefont {Lu}}, \ and\ \bibinfo {author}
  {\bibfnamefont {J.-W.}\ \bibnamefont {Pan}},\ }\bibfield  {title} {\bibinfo
  {title} {High-efficiency multiphoton boson sampling},\ }\href
  {\doibase/10.1038/nphoton.2017.63} {\bibfield  {journal} {\bibinfo  {journal}
  {Nature Photonics}\ }\textbf {\bibinfo {volume} {11}},\ \bibinfo {pages}
  {361-365} (\bibinfo {year} {2017})}\BibitemShut {NoStop}%
\bibitem [{\citenamefont {Agresti}\ \emph {et~al.}(2019)\citenamefont
  {Agresti}, \citenamefont {Viggianiello}, \citenamefont {Flamini},
  \citenamefont {Spagnolo}, \citenamefont {Crespi}, \citenamefont {Osellame},
  \citenamefont {Wiebe},\ and\ \citenamefont {Sciarrino}}]{Fabio2019}%
  \BibitemOpen
  \bibfield  {author} {\bibinfo {author} {\bibfnamefont {I.}~\bibnamefont
  {Agresti}}, \bibinfo {author} {\bibfnamefont {N.}~\bibnamefont
  {Viggianiello}}, \bibinfo {author} {\bibfnamefont {F.}~\bibnamefont
  {Flamini}}, \bibinfo {author} {\bibfnamefont {N.}~\bibnamefont {Spagnolo}},
  \bibinfo {author} {\bibfnamefont {A.}~\bibnamefont {Crespi}}, \bibinfo
  {author} {\bibfnamefont {R.}~\bibnamefont {Osellame}}, \bibinfo {author}
  {\bibfnamefont {N.}~\bibnamefont {Wiebe}}, \ and\ \bibinfo {author}
  {\bibfnamefont {F.}~\bibnamefont {Sciarrino}},\ }\bibfield  {title} {\bibinfo
  {title} {Pattern recognition techniques for boson sampling validation},\
  }\href {\doibase/10.1103/PhysRevX.9.011013} {\bibfield  {journal} {\bibinfo
  {journal} {Physical Review X}\ }\textbf {\bibinfo {volume} {9}},\ \bibinfo {pages}
  {011013} (\bibinfo {year} {2019})}\BibitemShut {NoStop}%
\bibitem [{\citenamefont {Zhong}\ \emph {et~al.}(2020)\citenamefont {Zhong},
  \citenamefont {Wang}, \citenamefont {Deng}, \citenamefont {Chen},
  \citenamefont {Peng}, \citenamefont {Luo}, \citenamefont {Qin}, \citenamefont
  {Wu}, \citenamefont {Ding}, \citenamefont {Hu}, \citenamefont {Hu},
  \citenamefont {Yang}, \citenamefont {Zhang}, \citenamefont {Li},
  \citenamefont {Li}, \citenamefont {Jiang}, \citenamefont {Gan}, \citenamefont
  {Yang}, \citenamefont {You}, \citenamefont {Wang}, \citenamefont {Li},
  \citenamefont {Liu}, \citenamefont {Lu},\ and\ \citenamefont
  {Pan}}]{Zhong2020}%
  \BibitemOpen
  \bibfield  {author} {\bibinfo {author} {\bibfnamefont {H.-S.}\ \bibnamefont
  {Zhong}}, \bibinfo {author} {\bibfnamefont {H.}~\bibnamefont {Wang}},
  \bibinfo {author} {\bibfnamefont {Y.-H.}\ \bibnamefont {Deng}}, \bibinfo
  {author} {\bibfnamefont {M.-C.}\ \bibnamefont {Chen}}, \bibinfo {author}
  {\bibfnamefont {L.-C.}\ \bibnamefont {Peng}}, \bibinfo {author}
  {\bibfnamefont {Y.-H.}\ \bibnamefont {Luo}}, \bibinfo {author} {\bibfnamefont
  {J.}~\bibnamefont {Qin}}, \bibinfo {author} {\bibfnamefont {D.}~\bibnamefont
  {Wu}}, \bibinfo {author} {\bibfnamefont {X.}~\bibnamefont {Ding}}, \bibinfo
  {author} {\bibfnamefont {Y.}~\bibnamefont {Hu}}, \bibinfo {author}
  {\bibfnamefont {P.}~\bibnamefont {Hu}}, \bibinfo {author} {\bibfnamefont
  {X.-Y.}\ \bibnamefont {Yang}}, \bibinfo {author} {\bibfnamefont {W.-J.}\
  \bibnamefont {Zhang}}, \bibinfo {author} {\bibfnamefont {H.}~\bibnamefont
  {Li}}, \bibinfo {author} {\bibfnamefont {Y.}~\bibnamefont {Li}}, \bibinfo
  {author} {\bibfnamefont {X.}~\bibnamefont {Jiang}}, \bibinfo {author}
  {\bibfnamefont {L.}~\bibnamefont {Gan}}, \bibinfo {author} {\bibfnamefont
  {G.}~\bibnamefont {Yang}}, \bibinfo {author} {\bibfnamefont {L.}~\bibnamefont
  {You}}, \bibinfo {author} {\bibfnamefont {Z.}~\bibnamefont {Wang}}, \bibinfo
  {author} {\bibfnamefont {L.}~\bibnamefont {Li}}, \bibinfo {author}
  {\bibfnamefont {N.-L.}\ \bibnamefont {Liu}}, \bibinfo {author} {\bibfnamefont
  {C.-Y.}\ \bibnamefont {Lu}}, \ and\ \bibinfo {author} {\bibfnamefont {J.-W.}\
  \bibnamefont {Pan}},\ }\bibfield  {title} {\bibinfo {title} {Quantum
  computational advantage using photons},\ }\href
  {\doibase/10.1126/science.abe8770} {\bibfield  {journal} {\bibinfo  {journal}
  {Science}\ }\textbf {\bibinfo {volume} {370}},\ \bibinfo {pages}
  {1460-1463} (\bibinfo {year} {2020})}\BibitemShut {NoStop}%
\bibitem [{\citenamefont {Simon}\ \emph {et~al.}(1994)\citenamefont {Simon},
  \citenamefont {Mukunda},\ and\ \citenamefont {Dutta}}]{Simon1994}%
  \BibitemOpen
  \bibfield  {author} {\bibinfo {author} {\bibfnamefont {R.}~\bibnamefont
  {Simon}}, \bibinfo {author} {\bibfnamefont {N.}~\bibnamefont {Mukunda}}, \
  and\ \bibinfo {author} {\bibfnamefont {B.}~\bibnamefont {Dutta}},\ }\bibfield
   {title} {\bibinfo {title} {Quantum-noise matrix for multimode systems: U(N)
  invariance, squeezing, and normal forms},\ }\href
  {\doibase/10.1103/physreva.49.1567} {\bibfield  {journal} {\bibinfo
  {journal} {Physical Review A}\ }\textbf {\bibinfo {volume} {49}},\ \bibinfo
  {pages} {1567} (\bibinfo {year} {1994})}\BibitemShut {NoStop}%
\bibitem [{\citenamefont {Arvind}\ \emph {et~al.}(1995)\citenamefont {Arvind},
  \citenamefont {Dutta}, \citenamefont {Mukunda},\ and\ \citenamefont
  {Simon}}]{Arvind1995}%
  \BibitemOpen
  \bibfield  {author} {\bibinfo {author} {\bibnamefont {Arvind}}, \bibinfo
  {author} {\bibfnamefont {B.}~\bibnamefont {Dutta}}, \bibinfo {author}
  {\bibfnamefont {N.}~\bibnamefont {Mukunda}}, \ and\ \bibinfo {author}
  {\bibfnamefont {R.}~\bibnamefont {Simon}},\ }\bibfield  {title} {\bibinfo
  {title} {The real symplectic groups in quantum mechanics and optics},\ }\href
  {\doibase/10.1007/bf02848172} {\bibfield  {journal} {\bibinfo  {journal}
  {Pramana}\ }\textbf {\bibinfo {volume} {45}},\ \bibinfo {pages} {471}
  (\bibinfo {year} {1995})}\BibitemShut {NoStop}%
\bibitem [{\citenamefont {de~Gosson}(2006)}]{deGosson2006}%
  \BibitemOpen
  \bibfield  {author} {\bibinfo {author} {\bibfnamefont {M.}~\bibnamefont
  {de~Gosson}},\ }\href {\doibase/10.1007/3-7643-7575-2} {\emph {\bibinfo
  {title} {Symplectic Geometry and Quantum Mechanics}}}\ (\bibinfo  {publisher}
  {Birkh\"{a}user Basel},\ \bibinfo {year} {2006})\BibitemShut {NoStop}%
\bibitem [{\citenamefont {Chatterjee}\ \emph {et~al.}(2026)\citenamefont {Chatterjee}, \citenamefont {Pandit}, \citenamefont {Singh}, \citenamefont {Chattopadhyay},\ and\ \citenamefont {Andersen}}]{Ulrik2026}%
\BibitemOpen
\bibfield  {author} {\bibinfo {author} {\bibfnamefont {K.}~\bibnamefont {Chatterjee}}, \bibinfo {author} {\bibfnamefont {T.}~\bibnamefont {Pandit}}, \bibinfo {author} {\bibfnamefont {V.}~\bibnamefont {Singh}}, \bibinfo {author} {\bibfnamefont {P.}~\bibnamefont {Chattopadhyay}}, \ and\ \bibinfo {author} {\bibfnamefont {U.~L.}\ \bibnamefont {Andersen}},\ } {\bibinfo {title} {Even odd splitting of the Gaussian quantum Fisher information: From symplectic geometry to metrology},\ } \href {\doibase/10.48550/ARXIV.2601.06513} {\bibinfo  {journal} {arXiv preprint arXiv:2601.06513}\ }(\bibinfo {year} {2026})\BibitemShut {NoStop}%
\bibitem [{\citenamefont {Wiseman}\ and\ \citenamefont
  {Milburn}(2009)}]{Milburn2009}%
  \BibitemOpen
  \bibfield  {author} {\bibinfo {author} {\bibfnamefont {H.~M.}\ \bibnamefont
  {Wiseman}}\ and\ \bibinfo {author} {\bibfnamefont {G.~J.}\ \bibnamefont
  {Milburn}},\ }\href@noop {} {\emph {\bibinfo {title} {Quantum Measurement and
  Control}}}\ (\bibinfo  {publisher} {Cambridge University Press},\ \bibinfo
  {year} {2009})\BibitemShut {NoStop}%
\bibitem [{\citenamefont {Jing}\ \emph {et~al.}(2011)\citenamefont {Jing},
  \citenamefont {Liu}, \citenamefont {Zhou}, \citenamefont {Ou},\ and\
  \citenamefont {Zhang}}]{Jing2011}%
  \BibitemOpen
  \bibfield  {author} {\bibinfo {author} {\bibfnamefont {J.}~\bibnamefont
  {Jing}}, \bibinfo {author} {\bibfnamefont {C.}~\bibnamefont {Liu}}, \bibinfo
  {author} {\bibfnamefont {Z.}~\bibnamefont {Zhou}}, \bibinfo {author}
  {\bibfnamefont {Z.~Y.}\ \bibnamefont {Ou}}, \ and\ \bibinfo {author}
  {\bibfnamefont {W.}~\bibnamefont {Zhang}},\ }\bibfield  {title} {\bibinfo
  {title} {Realization of a nonlinear interferometer with parametric
  amplifiers},\ }\href {\doibase/10.1063/1.3606549} {\bibfield  {journal}
  {\bibinfo  {journal} {Applied Physics Letters}\ }\textbf {\bibinfo {volume}
  {99}},\ \bibinfo {pages} {011110} (\bibinfo {year} {2011})}\BibitemShut {NoStop}%
\bibitem [{\citenamefont {Hudelist}\ \emph {et~al.}(2014)\citenamefont
  {Hudelist}, \citenamefont {Kong}, \citenamefont {Liu}, \citenamefont {Jing},
  \citenamefont {Ou},\ and\ \citenamefont {Zhang}}]{Hudelist2014}%
  \BibitemOpen
  \bibfield  {author} {\bibinfo {author} {\bibfnamefont {F.}~\bibnamefont
  {Hudelist}}, \bibinfo {author} {\bibfnamefont {J.}~\bibnamefont {Kong}},
  \bibinfo {author} {\bibfnamefont {C.}~\bibnamefont {Liu}}, \bibinfo {author}
  {\bibfnamefont {J.}~\bibnamefont {Jing}}, \bibinfo {author} {\bibfnamefont
  {Z.}~\bibnamefont {Ou}}, \ and\ \bibinfo {author} {\bibfnamefont
  {W.}~\bibnamefont {Zhang}},\ }\bibfield  {title} {\bibinfo {title} {Quantum
  metrology with parametric amplifier-based photon correlation
  interferometers},\ }\href {\doibase/10.1038/ncomms4049} {\bibfield  {journal}
  {\bibinfo  {journal} {Nature Communications}\ }\textbf {\bibinfo {volume}
  {5}},\ \bibinfo {pages} {3049} (\bibinfo {year} {2014})}\BibitemShut {NoStop}%
\bibitem [{\citenamefont {Chen}\ \emph {et~al.}(2015)\citenamefont {Chen},
  \citenamefont {Qiu}, \citenamefont {Chen}, \citenamefont {Guo}, \citenamefont
  {Chen}, \citenamefont {Ou},\ and\ \citenamefont {Zhang}}]{Zhang2015}%
  \BibitemOpen
  \bibfield  {author} {\bibinfo {author} {\bibfnamefont {B.}~\bibnamefont
  {Chen}}, \bibinfo {author} {\bibfnamefont {C.}~\bibnamefont {Qiu}}, \bibinfo
  {author} {\bibfnamefont {S.}~\bibnamefont {Chen}}, \bibinfo {author}
  {\bibfnamefont {J.}~\bibnamefont {Guo}}, \bibinfo {author} {\bibfnamefont
  {L.~Q.}\ \bibnamefont {Chen}}, \bibinfo {author} {\bibfnamefont {Z.~Y.}\
  \bibnamefont {Ou}}, \ and\ \bibinfo {author} {\bibfnamefont {W.}~\bibnamefont
  {Zhang}},\ }\bibfield  {title} {\bibinfo {title} {Atom-light hybrid
  interferometer},\ }\href {\doibase/10.1103/PhysRevLett.115.043602} {\bibfield
   {journal} {\bibinfo  {journal} {Physical Review Letters}\ }\textbf {\bibinfo
  {volume} {115}},\ \bibinfo {pages} {043602} (\bibinfo {year}
  {2015})}\BibitemShut {NoStop}%
\bibitem [{\citenamefont {Gabbrielli}\ \emph {et~al.}(2015)\citenamefont
  {Gabbrielli}, \citenamefont {Pezz\`e},\ and\ \citenamefont
  {Smerzi}}]{Augusto2015}%
  \BibitemOpen
  \bibfield  {author} {\bibinfo {author} {\bibfnamefont {M.}~\bibnamefont
  {Gabbrielli}}, \bibinfo {author} {\bibfnamefont {L.}~\bibnamefont {Pezz\`e}},
  \ and\ \bibinfo {author} {\bibfnamefont {A.}~\bibnamefont {Smerzi}},\
  }\bibfield  {title} {\bibinfo {title} {Spin-mixing interferometry with
  Bose-Einstein condensates},\ }\href {\doibase/10.1103/PhysRevLett.115.163002}
  {\bibfield  {journal} {\bibinfo  {journal} {Physical Review Letters}\ }\textbf
  {\bibinfo {volume} {115}},\ \bibinfo {pages} {163002} (\bibinfo {year}
  {2015})}\BibitemShut {NoStop}%
\bibitem [{\citenamefont {Linnemann}\ \emph {et~al.}(2016)\citenamefont
  {Linnemann}, \citenamefont {Strobel}, \citenamefont {Muessel}, \citenamefont
  {Schulz}, \citenamefont {Lewis-Swan}, \citenamefont {Kheruntsyan},\ and\
  \citenamefont {Oberthaler}}]{Oberthaler2016}%
  \BibitemOpen
  \bibfield  {author} {\bibinfo {author} {\bibfnamefont {D.}~\bibnamefont
  {Linnemann}}, \bibinfo {author} {\bibfnamefont {H.}~\bibnamefont {Strobel}},
  \bibinfo {author} {\bibfnamefont {W.}~\bibnamefont {Muessel}}, \bibinfo
  {author} {\bibfnamefont {J.}~\bibnamefont {Schulz}}, \bibinfo {author}
  {\bibfnamefont {R.~J.}\ \bibnamefont {Lewis-Swan}}, \bibinfo {author}
  {\bibfnamefont {K.~V.}\ \bibnamefont {Kheruntsyan}}, \ and\ \bibinfo {author}
  {\bibfnamefont {M.~K.}\ \bibnamefont {Oberthaler}},\ }\bibfield  {title}
  {\bibinfo {title} {Quantum-enhanced sensing based on time reversal of
  nonlinear dynamics},\ }\href {\doibase/10.1103/PhysRevLett.117.013001}
  {\bibfield  {journal} {\bibinfo  {journal} {Physical Review Letters}\ }\textbf
  {\bibinfo {volume} {117}},\ \bibinfo {pages} {013001} (\bibinfo {year}
  {2016})}\BibitemShut {NoStop}%
\bibitem [{\citenamefont {Szigeti}\ \emph {et~al.}(2017)\citenamefont
  {Szigeti}, \citenamefont {Lewis-Swan},\ and\ \citenamefont
  {Haine}}]{Haine2017}%
  \BibitemOpen
  \bibfield  {author} {\bibinfo {author} {\bibfnamefont {S.~S.}\ \bibnamefont
  {Szigeti}}, \bibinfo {author} {\bibfnamefont {R.~J.}\ \bibnamefont
  {Lewis-Swan}}, \ and\ \bibinfo {author} {\bibfnamefont {S.~A.}\ \bibnamefont
  {Haine}},\ }\bibfield  {title} {\bibinfo {title} {Pumped-up SU(1,1)
  interferometry},\ }\href {\doibase/10.1103/PhysRevLett.118.150401} {\bibfield
   {journal} {\bibinfo  {journal} {Physical Review Letters}\ }\textbf {\bibinfo
  {volume} {118}},\ \bibinfo {pages} {150401} (\bibinfo {year}
  {2017})}\BibitemShut {NoStop}%
\bibitem [{\citenamefont {Anderson}\ \emph {et~al.}(2017)\citenamefont
  {Anderson}, \citenamefont {Gupta}, \citenamefont {Schmittberger},
  \citenamefont {Horrom}, \citenamefont {Hermann-Avigliano}, \citenamefont
  {Jones},\ and\ \citenamefont {Lett}}]{Lett2017}%
  \BibitemOpen
  \bibfield  {author} {\bibinfo {author} {\bibfnamefont {B.~E.}\ \bibnamefont
  {Anderson}}, \bibinfo {author} {\bibfnamefont {P.}~\bibnamefont {Gupta}},
  \bibinfo {author} {\bibfnamefont {B.~L.}\ \bibnamefont {Schmittberger}},
  \bibinfo {author} {\bibfnamefont {T.}~\bibnamefont {Horrom}}, \bibinfo
  {author} {\bibfnamefont {C.}~\bibnamefont {Hermann-Avigliano}}, \bibinfo
  {author} {\bibfnamefont {K.~M.}\ \bibnamefont {Jones}}, \ and\ \bibinfo
  {author} {\bibfnamefont {P.~D.}\ \bibnamefont {Lett}},\ }\bibfield  {title}
  {\bibinfo {title} {Phase sensing beyond the standard quantum limit with a
  variation on the SU(1,1) interferometer},\ }\href
  {\doibase/10.1364/OPTICA.4.000752} {\bibfield  {journal} {\bibinfo  {journal}
  {Optica}\ }\textbf {\bibinfo {volume} {4}},\ \bibinfo {pages} {752} (\bibinfo
  {year} {2017})}\BibitemShut {NoStop}%
\bibitem [{\citenamefont {Manceau}\ \emph {et~al.}(2017)\citenamefont
  {Manceau}, \citenamefont {Leuchs}, \citenamefont {Khalili},\ and\
  \citenamefont {Chekhova}}]{Maria2017}%
  \BibitemOpen
  \bibfield  {author} {\bibinfo {author} {\bibfnamefont {M.}~\bibnamefont
  {Manceau}}, \bibinfo {author} {\bibfnamefont {G.}~\bibnamefont {Leuchs}},
  \bibinfo {author} {\bibfnamefont {F.}~\bibnamefont {Khalili}}, \ and\
  \bibinfo {author} {\bibfnamefont {M.}~\bibnamefont {Chekhova}},\ }\bibfield
  {title} {\bibinfo {title} {Detection loss tolerant supersensitive phase
  measurement with an SU(1,1) interferometer},\ }\href
  {\doibase/10.1103/PhysRevLett.119.223604} {\bibfield  {journal} {\bibinfo
  {journal} {Physical Review Letters}\ }\textbf {\bibinfo {volume} {119}},\ \bibinfo
  {pages} {223604} (\bibinfo {year} {2017})}\BibitemShut {NoStop}%
\bibitem [{\citenamefont {Lukens}\ \emph {et~al.}(2018)\citenamefont {Lukens},
  \citenamefont {Pooser},\ and\ \citenamefont {Peters}}]{Lukens2018}%
  \BibitemOpen
  \bibfield  {author} {\bibinfo {author} {\bibfnamefont {J.~M.}\ \bibnamefont
  {Lukens}}, \bibinfo {author} {\bibfnamefont {R.~C.}\ \bibnamefont {Pooser}},
  \ and\ \bibinfo {author} {\bibfnamefont {N.~A.}\ \bibnamefont {Peters}},\
  }\bibfield  {title} {\bibinfo {title} {A broadband fiber-optic nonlinear
  interferometer},\ }\href {\doibase/10.1063/1.5048198} {\bibfield  {journal}
  {\bibinfo  {journal} {Applied Physics Letters}\ }\textbf {\bibinfo {volume}
  {113}},\ \bibinfo {pages} {091103} (\bibinfo {year} {2018})}\BibitemShut {NoStop}%
\bibitem [{\citenamefont {Pezzè}\ and\ \citenamefont {Smerzi}(2014)}]{Smerzi2014}%
  \BibitemOpen
  \bibfield  {author} {\bibinfo {author} {\bibfnamefont {L.}~\bibnamefont
  {Pezzè}}\ and\ \bibinfo {author} {\bibfnamefont {A.}~\bibnamefont {Smerzi}},\
  }\bibfield  {title} {\bibinfo {title} {Quantum theory of phase estimation},\
  }\href {\doibase/10.3254/978-1-61499-448-0-691} {\bibfield  {journal}
  {\bibinfo  {journal} {Proceedings of the International School of Physics
  ``Enrico Fermi''}\ }\textbf {\bibinfo {volume} {188}},\ \bibinfo {pages}
  {691} (\bibinfo {year} {2014})}\BibitemShut {NoStop}%
\bibitem [{\citenamefont {Peskin}(2018)}]{Peskin2018}%
  \BibitemOpen
  \bibfield  {author} {\bibinfo {author} {\bibfnamefont {M.~E.}\ \bibnamefont
  {Peskin}},\ }\href {\doibase/10.1201/9780429503559} {\emph {\bibinfo {title}
  {An Introduction To Quantum Field Theory}}}\ (\bibinfo  {publisher} {CRC
  Press},\ \bibinfo {year} {2018})\BibitemShut {NoStop}%
\bibitem [{\citenamefont {Nielsen}\ \emph {et~al.}(2006)\citenamefont
  {Nielsen}, \citenamefont {Dowling}, \citenamefont {Gu},\ and\ \citenamefont
  {Doherty}}]{Nielsen2006}%
  \BibitemOpen
  \bibfield  {author} {\bibinfo {author} {\bibfnamefont {M.~A.}\ \bibnamefont
  {Nielsen}}, \bibinfo {author} {\bibfnamefont {M.~R.}\ \bibnamefont
  {Dowling}}, \bibinfo {author} {\bibfnamefont {M.}~\bibnamefont {Gu}}, \ and\
  \bibinfo {author} {\bibfnamefont {A.~C.}\ \bibnamefont {Doherty}},\
  }\bibfield  {title} {\bibinfo {title} {Quantum computation as geometry},\
  }\href {\doibase/10.1126/science.1121541} {\bibfield  {journal} {\bibinfo
  {journal} {Science}\ }\textbf {\bibinfo {volume} {311}},\ \bibinfo {pages}
  {1133-1135} (\bibinfo {year} {2006})}\BibitemShut {NoStop}%
\bibitem [{\citenamefont {Provost}\ and\ \citenamefont
  {Vallee}(1980)}]{Provost1980}%
  \BibitemOpen
  \bibfield  {author} {\bibinfo {author} {\bibfnamefont {J.~P.}\ \bibnamefont
  {Provost}}\ and\ \bibinfo {author} {\bibfnamefont {G.}~\bibnamefont
  {Vallee}},\ }\bibfield  {title} {\bibinfo {title} {Riemannian structure on
  manifolds of quantum states},\ }\href {\doibase/10.1007/bf02193559}
  {\bibfield  {journal} {\bibinfo  {journal} {Communications in Mathematical
  Physics}\ }\textbf {\bibinfo {volume} {76}},\ \bibinfo {pages} {289-301}
  (\bibinfo {year} {1980})}\BibitemShut {NoStop}%
\bibitem [{\citenamefont {Braunstein}\ and\ \citenamefont
  {Caves}(1994)}]{Caves1994}%
  \BibitemOpen
  \bibfield  {author} {\bibinfo {author} {\bibfnamefont {S.~L.}\ \bibnamefont
  {Braunstein}}\ and\ \bibinfo {author} {\bibfnamefont {C.~M.}\ \bibnamefont
  {Caves}},\ }\bibfield  {title} {\bibinfo {title} {Statistical distance and
  the geometry of quantum states},\ }\href
  {\doibase/10.1103/PhysRevLett.72.3439} {\bibfield  {journal} {\bibinfo
  {journal} {Physical Review Letters}\ }\textbf {\bibinfo {volume} {72}},\ \bibinfo
  {pages} {3439} (\bibinfo {year} {1994})}\BibitemShut {NoStop}%
\bibitem [{\citenamefont {Braunstein}(2005)}]{Braunstein2005_BM}%
  \BibitemOpen
  \bibfield  {author} {\bibinfo {author} {\bibfnamefont {S.~L.}\ \bibnamefont
  {Braunstein}},\ }\bibfield  {title} {\bibinfo {title} {Squeezing as an
  irreducible resource},\ }\href {\doibase/10.1103/physreva.71.055801}
  {\bibfield  {journal} {\bibinfo  {journal} {Physical Review A}\ }\textbf
  {\bibinfo {volume} {71}},\ \bibinfo
  {pages} {055801} (\bibinfo {year} {2005})}\BibitemShut {NoStop}%
\bibitem [{\citenamefont {Houde}\ \emph {et~al.}(2024)\citenamefont {Houde},
  \citenamefont {McCutcheon},\ and\ \citenamefont {Quesada}}]{Houde2024}%
  \BibitemOpen
  \bibfield  {author} {\bibinfo {author} {\bibfnamefont {M.}~\bibnamefont
  {Houde}}, \bibinfo {author} {\bibfnamefont {W.}~\bibnamefont {McCutcheon}}, \
  and\ \bibinfo {author} {\bibfnamefont {N.}~\bibnamefont {Quesada}},\
  }\bibfield  {title} {\bibinfo {title} {Matrix decompositions in quantum
  optics: Takagi/Autonne, Bloch-Messiah/Euler, Iwasawa, and Williamson},\
  }\href {\doibase/10.1139/cjp-2024-0070} {\bibfield  {journal} {\bibinfo
  {journal} {Canadian Journal of Physics}\ }\textbf {\bibinfo {volume} {102}},\
  \bibinfo {pages} {497-507} (\bibinfo {year} {2024})}\BibitemShut {NoStop}%
\bibitem [{\citenamefont {Nielsen}(2005)}]{Nielsen2005}%
  \BibitemOpen
  \bibfield  {author} {\bibinfo {author} {\bibfnamefont {M.~A.}\ \bibnamefont
  {Nielsen}},\ }{\bibinfo
  {title} {A geometric approach to quantum circuit lower bounds},\ }\href {\doibase/10.48550/ARXIV.QUANT-PH/0502070} {\bibinfo
  {journal} {arXiv preprint quant-ph/0502070}\ } (\bibinfo
  {year} {2005})\BibitemShut {NoStop}%
\bibitem [{\citenamefont {Pfeifer}(2019)}]{Pfeifer2019}%
  \BibitemOpen
  \bibfield  {author} {\bibinfo {author} {\bibfnamefont {C.}~\bibnamefont
  {Pfeifer}},\ }\bibfield  {title} {\bibinfo {title} {Finsler spacetime
  geometry in physics},\ }\href {\doibase/10.1142/s0219887819410044} {\bibfield
   {journal} {\bibinfo  {journal} {International Journal of Geometric Methods
  in Modern Physics}\ }\textbf {\bibinfo {volume} {16}},\ \bibinfo {pages}
  {1941004} (\bibinfo {year} {2019})}\BibitemShut {NoStop}%
\bibitem [{\citenamefont {Gardiner}\ and\ \citenamefont
  {Zoller}(2004)}]{Crispin2004}%
  \BibitemOpen
  \bibfield  {author} {\bibinfo {author} {\bibfnamefont {C.}~\bibnamefont
  {Gardiner}}\ and\ \bibinfo {author} {\bibfnamefont {P.}~\bibnamefont
  {Zoller}},\ }\href@noop {} {\emph {\bibinfo {title} {Quantum noise: a
  handbook of Markovian and non-Markovian quantum stochastic methods with
  applications to quantum optics}}},\ \bibinfo {edition} {3rd}\ ed.,\ Springer
  series in synergetics\ (\bibinfo  {publisher} {Springer},\ \bibinfo {address}
  {Springer Berlin, Heidelberg},\ \bibinfo {year} {2004})\BibitemShut {NoStop}%
\bibitem [{\citenamefont {Chen}\ \emph {et~al.}(2020)\citenamefont {Chen},
  \citenamefont {Zhang}, \citenamefont {Zheng}, \citenamefont {Wu},\ and\
  \citenamefont {Zhai}}]{Chen2020}%
  \BibitemOpen
  \bibfield  {author} {\bibinfo {author} {\bibfnamefont {Y.-Y.}\ \bibnamefont
  {Chen}}, \bibinfo {author} {\bibfnamefont {P.}~\bibnamefont {Zhang}},
  \bibinfo {author} {\bibfnamefont {W.}~\bibnamefont {Zheng}}, \bibinfo
  {author} {\bibfnamefont {Z.}~\bibnamefont {Wu}}, \ and\ \bibinfo {author}
  {\bibfnamefont {H.}~\bibnamefont {Zhai}},\ }\bibfield  {title} {\bibinfo
  {title} {Many-body echo},\ }\href {\doibase/10.1103/PhysRevA.102.011301}
  {\bibfield  {journal} {\bibinfo  {journal} {Physical Review A}\ }\textbf {\bibinfo
  {volume} {102}},\ \bibinfo {pages} {011301} (\bibinfo {year}
  {2020})}\BibitemShut {NoStop}%
\bibitem [{\citenamefont {Lyu}\ \emph {et~al.}(2020)\citenamefont {Lyu},
  \citenamefont {Lv},\ and\ \citenamefont {Zhou}}]{Lyu2020}%
  \BibitemOpen
  \bibfield  {author} {\bibinfo {author} {\bibfnamefont {C.}~\bibnamefont
  {Lyu}}, \bibinfo {author} {\bibfnamefont {C.}~\bibnamefont {Lv}}, \ and\
  \bibinfo {author} {\bibfnamefont {Q.}~\bibnamefont {Zhou}},\ }\bibfield
  {title} {\bibinfo {title} {Geometrizing quantum dynamics of a Bose-Einstein
  condensate},\ }\href {\doibase/10.1103/physrevlett.125.253401} {\bibfield
  {journal} {\bibinfo  {journal} {Physical Review Letters}\ }\textbf {\bibinfo
  {volume} {125}},\ \bibinfo {pages} {253401} (\bibinfo {year} {2020})}\BibitemShut {NoStop}%
\bibitem [{\citenamefont {Lv}\ \emph {et~al.}(2020)\citenamefont {Lv},
  \citenamefont {Zhang},\ and\ \citenamefont {Zhou}}]{Lv2020}%
  \BibitemOpen
  \bibfield  {author} {\bibinfo {author} {\bibfnamefont {C.}~\bibnamefont
  {Lv}}, \bibinfo {author} {\bibfnamefont {R.}~\bibnamefont {Zhang}}, \ and\
  \bibinfo {author} {\bibfnamefont {Q.}~\bibnamefont {Zhou}},\ }\bibfield
  {title} {\bibinfo {title} {$su(1,1)$ echoes for breathers in quantum gases},\
  }\href {\doibase/10.1103/PhysRevLett.125.253002} {\bibfield  {journal}
  {\bibinfo  {journal} {Physical Review Letters}\ }\textbf {\bibinfo {volume} {125}},\ \bibinfo {pages} {253002} (\bibinfo {year} {2020})}\BibitemShut {NoStop}%
\bibitem [{\citenamefont {Wang}(2024)}]{Wang2024}%
  \BibitemOpen
  \bibfield  {author} {\bibinfo {author} {\bibfnamefont {C.-Y.}\ \bibnamefont
  {Wang}},\ }\bibfield  {title} {\bibinfo {title} {Quantum echo in
  two-component Bose-Einstein condensates},\ }\href
  {\doibase/10.1103/physreva.109.063327} {\bibfield  {journal} {\bibinfo
  {journal} {Physical Review A}\ }\textbf {\bibinfo {volume} {109}},\ \bibinfo {pages} {063327} (\bibinfo
  {year} {2024})}\BibitemShut {NoStop}%
\bibitem [{\citenamefont {Wisniacki}(2012)}]{Wisniacki2012}%
  \BibitemOpen
  \bibfield  {author} {\bibinfo {author} {\bibfnamefont {A.}~\bibnamefont
  {Wisniacki}},\ }\bibfield  {title} {\bibinfo {title} {Loschmidt echo},\
  }\href {\doibase/10.4249/scholarpedia.11687} {\bibfield  {journal} {\bibinfo
  {journal} {Scholarpedia}\ }\textbf {\bibinfo {volume} {7}},\ \bibinfo {pages}
  {11687} (\bibinfo {year} {2012})}\BibitemShut {NoStop}%
\bibitem [{\citenamefont {Macr\`{\i}}\ \emph {et~al.}(2016)\citenamefont
  {Macr\`{\i}}, \citenamefont {Smerzi},\ and\ \citenamefont
  {Pezz\`e}}]{Tommaso2016}%
  \BibitemOpen
  \bibfield  {author} {\bibinfo {author} {\bibfnamefont {T.}~\bibnamefont
  {Macr\`{\i}}}, \bibinfo {author} {\bibfnamefont {A.}~\bibnamefont {Smerzi}},
  \ and\ \bibinfo {author} {\bibfnamefont {L.}~\bibnamefont {Pezz\`e}},\
  }\bibfield  {title} {\bibinfo {title} {Loschmidt echo for quantum
  metrology},\ }\href {\doibase/10.1103/PhysRevA.94.010102} {\bibfield
  {journal} {\bibinfo  {journal} {Physical Review A}\ }\textbf {\bibinfo {volume}
  {94}},\ \bibinfo {pages} {010102} (\bibinfo {year} {2016})}\BibitemShut
  {NoStop}%
\bibitem [{\citenamefont {Fabre}\ and\ \citenamefont
  {Treps}(2020)}]{Fabre2020}%
  \BibitemOpen
  \bibfield  {author} {\bibinfo {author} {\bibfnamefont {C.}~\bibnamefont
  {Fabre}}\ and\ \bibinfo {author} {\bibfnamefont {N.}~\bibnamefont {Treps}},\
  }\bibfield  {title} {\bibinfo {title} {Modes and states in quantum optics},\
  }\href {\doibase/10.1103/revmodphys.92.035005} {\bibfield  {journal}
  {\bibinfo  {journal} {Reviews of Modern Physics}\ }\textbf {\bibinfo {volume}
  {92}},\ \bibinfo {pages} {035005} (\bibinfo {year} {2020})}\BibitemShut
  {NoStop}%
\bibitem [{\citenamefont {McDonald}\ \emph {et~al.}(2018)\citenamefont
  {McDonald}, \citenamefont {Pereg-Barnea},\ and\ \citenamefont
  {Clerk}}]{Clerk2018}%
  \BibitemOpen
  \bibfield  {author} {\bibinfo {author} {\bibfnamefont {A.}~\bibnamefont
  {McDonald}}, \bibinfo {author} {\bibfnamefont {T.}~\bibnamefont
  {Pereg-Barnea}}, \ and\ \bibinfo {author} {\bibfnamefont {A.~A.}\
  \bibnamefont {Clerk}},\ }\bibfield  {title} {\bibinfo {title}
  {Phase-dependent chiral transport and effective non-Hermitian dynamics in a
  bosonic Kitaev-Majorana chain},\ }\href {\doibase/10.1103/PhysRevX.8.041031}
  {\bibfield  {journal} {\bibinfo  {journal} {Physical Review X}\ }\textbf {\bibinfo
  {volume} {8}},\ \bibinfo {pages} {041031} (\bibinfo {year}
  {2018})}\BibitemShut {NoStop}%
\bibitem [{\citenamefont {Lv}\ and\ \citenamefont {Zhou}(2024)}]{LV2024}%
  \BibitemOpen
  \bibfield  {author} {\bibinfo {author} {\bibfnamefont {C.}~\bibnamefont
  {Lv}}\ and\ \bibinfo {author} {\bibfnamefont {Q.}~\bibnamefont {Zhou}},\
  } {\bibinfo {title} {Hidden curved
  spaces in bosonic Kitaev model},\ }\href {\doibase/10.48550/ARXIV.2408.05132} {\bibinfo  {journal} {arXiv preprint arXiv:2408.05132}\ }(\bibinfo {year} {2024})\BibitemShut
  {NoStop}%
\bibitem [{\citenamefont {Lee}\ \emph {et~al.}(2024)\citenamefont {Lee},
  \citenamefont {Jin}, \citenamefont {Wang}, \citenamefont {McDonald},\ and\
  \citenamefont {Clerk}}]{Lee2024}%
  \BibitemOpen
  \bibfield  {author} {\bibinfo {author} {\bibfnamefont {G.}~\bibnamefont
  {Lee}}, \bibinfo {author} {\bibfnamefont {T.}~\bibnamefont {Jin}}, \bibinfo
  {author} {\bibfnamefont {Y.-X.}\ \bibnamefont {Wang}}, \bibinfo {author}
  {\bibfnamefont {A.}~\bibnamefont {McDonald}}, \ and\ \bibinfo {author}
  {\bibfnamefont {A.}~\bibnamefont {Clerk}},\ }\bibfield  {title} {\bibinfo
  {title} {Entanglement phase transition due to reciprocity breaking without
  measurement or postselection},\ }\href {\doibase/10.1103/prxquantum.5.010313}
  {\bibfield  {journal} {\bibinfo  {journal} {PRX Quantum}\ }\textbf {\bibinfo
  {volume} {5}},\ \bibinfo {pages} {010313} (\bibinfo {year} {2024})}\BibitemShut {NoStop}%
\bibitem [{\citenamefont {Hatano}\ and\ \citenamefont
  {Nelson}(1996)}]{Hatano1996}%
  \BibitemOpen
  \bibfield  {author} {\bibinfo {author} {\bibfnamefont {N.}~\bibnamefont
  {Hatano}}\ and\ \bibinfo {author} {\bibfnamefont {D.~R.}\ \bibnamefont
  {Nelson}},\ }\bibfield  {title} {\bibinfo {title} {Localization transitions
  in non-Hermitian quantum mechanics},\ }\href
  {\doibase/10.1103/physrevlett.77.570} {\bibfield  {journal} {\bibinfo
  {journal} {Physical Review Letters}\ }\textbf {\bibinfo {volume} {77}},\
  \bibinfo {pages} {570} (\bibinfo {year} {1996})}\BibitemShut {NoStop}%
\end{thebibliography}

\begin{thebibliography}{70}%
\makeatletter
\providecommand \@ifxundefined [1]{%
 \@ifx{#1\undefined}
}%
\providecommand \@ifnum [1]{%
 \ifnum #1\expandafter \@firstoftwo
 \else \expandafter \@secondoftwo
 \fi
}%
\providecommand \@ifx [1]{%
 \ifx #1\expandafter \@firstoftwo
 \else \expandafter \@secondoftwo
 \fi
}%
\providecommand \natexlab [1]{#1}%
\providecommand \enquote  [1]{#1}%
\providecommand \bibnamefont  [1]{#1}%
\providecommand \bibfnamefont [1]{#1}%
\providecommand \citenamefont [1]{#1}%
\providecommand \href@noop [0]{\@secondoftwo}%
\providecommand \href [0]{\begingroup \@sanitize@url \@href}%
\providecommand \@href[1]{\@@startlink{#1}\@@href}%
\providecommand \@@href[1]{\endgroup#1\@@endlink}%
\providecommand \@sanitize@url [0]{\catcode `\\12\catcode `\$12\catcode
  `\&12\catcode `\#12\catcode `\^12\catcode `\_12\catcode `\%12\relax}%
\providecommand \@@startlink[1]{}%
\providecommand \@@endlink[0]{}%
\providecommand \url  [0]{\begingroup\@sanitize@url \@url }%
\providecommand \@url [1]{\endgroup\@href {#1}{\urlprefix }}%
\providecommand \urlprefix  [0]{URL }%
\providecommand \Eprint [0]{\href }%
\providecommand \doibase [0]{https://dx.doi.org}%
\providecommand \selectlanguage [0]{\@gobble}%
\providecommand \bibinfo  [0]{\@secondoftwo}%
\providecommand \bibfield  [0]{\@secondoftwo}%
\providecommand \translation [1]{[#1]}%
\providecommand \BibitemOpen [0]{}%
\providecommand \bibitemStop [0]{}%
\providecommand \bibitemNoStop [0]{.\EOS\space}%
\providecommand \EOS [0]{\spacefactor3000\relax}%
\providecommand \BibitemShut  [1]{\csname bibitem#1\endcsname}%
\let\auto@bib@innerbib\@empty
\bibitem [{\citenamefont {Pezzè}\ and\ \citenamefont {Smerzi}(2014)}]{Smerzi2014S}%
  \BibitemOpen
  \bibfield  {author} {\bibinfo {author} {\bibfnamefont {L.}~\bibnamefont
  {Pezzè}}\ and\ \bibinfo {author} {\bibfnamefont {A.}~\bibnamefont {Smerzi}},\
  }\bibfield  {title} {\bibinfo {title} {Quantum theory of phase estimation},\
  }\href {\doibase/10.3254/978-1-61499-448-0-691} {\bibfield  {journal}
  {\bibinfo  {journal} {Proceedings of the International School of Physics
  ``Enrico Fermi''}\ }\textbf {\bibinfo {volume} {188}},\ \bibinfo {pages}
  {691} (\bibinfo {year} {2014})}\BibitemShut {NoStop}%
\end{thebibliography}
\end{document}